\documentclass{article}

\usepackage{microtype}
\usepackage{graphicx}
\usepackage{subfigure}
\usepackage{booktabs} 

\usepackage{amsmath,amsfonts,bm}

\def\eqref#1{equation~\ref{#1}}

\def\1{\bm{1}}

\DeclareMathAlphabet{\mathsfit}{\encodingdefault}{\sfdefault}{m}{sl}
\SetMathAlphabet{\mathsfit}{bold}{\encodingdefault}{\sfdefault}{bx}{n}

\usepackage{hyperref}
\usepackage{multicol}
\usepackage{multirow}

\usepackage{wrapfig}
\usepackage[accepted]{icml2024}

\usepackage{hyperref}
\usepackage{amsmath}
\usepackage{amssymb}
\usepackage{mathtools}
\usepackage{amsthm}
\usepackage{caption}
\usepackage{xcolor}
\usepackage[capitalize,noabbrev]{cleveref}
\usepackage{arydshln} 
\usepackage[most]{tcolorbox}
\theoremstyle{plain}
\newtheorem{theorem}{Theorem}[section]
\newtheorem{proposition}[theorem]{Proposition}

\theoremstyle{definition}

\theoremstyle{remark}

\usepackage[textsize=tiny]{todonotes}

\icmltitlerunning{\textsc{PPFlow}: Target-aware Peptide Design with Torsional Flow Matching}

\begin{document}

\twocolumn[
\icmltitle{\textsc{PPFlow}: Target-Aware Peptide Design with Torsional Flow Matching}



\icmlsetsymbol{equal}{*}

\begin{icmlauthorlist}
\icmlauthor{Haitao Lin}{zju,wlu}
\icmlauthor{Odin Zhang}{zju}
\icmlauthor{Huifeng Zhao}{zju}
\icmlauthor{Dejun Jiang}{zju} \\
\icmlauthor{Lirong Wu}{wlu,zju}
\icmlauthor{Zicheng Liu}{wlu,zju}
\icmlauthor{Yufei Huang}{wlu,zju}
\icmlauthor{Stan Z. Li$^\dagger$}{wlu}
\end{icmlauthorlist}

\icmlaffiliation{zju}{Zhejiang University;}
\icmlaffiliation{wlu}{AI Lab, Research Center for Industries of the Future, Westlake University, Hangzhou, China}

\icmlcorrespondingauthor{Stan Z. Li}{stan.zq.li@westlake.edu.cn}


\vskip 0.3in
]


\printAffiliationsAndNotice{}

\begin{abstract}
Therapeutic peptides have proven to have great pharmaceutical value and potential in recent decades. However, methods of AI-assisted peptide drug discovery are not fully explored.
To fill the gap, we propose a target-aware peptide design method called \textsc{PPFlow}, based on conditional flow matching on torus manifolds, to model the internal geometries of torsion angles for the peptide structure design. Besides, we establish a protein-peptide binding dataset named \texttt{PPBench2024} to fill the void of massive data for the task of structure-based peptide drug design and to allow the training of deep learning methods. Extensive experiments show that \textsc{PPFlow} reaches state-of-the-art performance in tasks of peptide drug generation and optimization in comparison with baseline models, and can be generalized to other tasks including docking and side-chain packing.
\end{abstract}
\vspace{-2em}
\section{Introduction}
\label{introduction}
Therapeutic peptides are a unique class of pharmaceutical agents composed of a series of well-ordered amino acids. The development of peptide drug design and discovery is accelerated by fast advances in structural biology, recombinant biologics, and synthetic and analytic technologies \cite{Wang2022TherapeuticPC}. Peptide drugs play an important role in pharmacology because they usually bind to cell surface receptors and trigger intracellular effects with high affinity and specificity, showing less immunogenicity and taking lower production costs \cite{Muttenthaler2021TrendsIP}. Therefore, they have raised great research interests as new peptide therapeutics are continuously developed with more than 150 peptides in clinical tests and another 400–600 peptides undergoing preclinical studies.

Deep learning has revolutionized fields like drug discovery and protein design \cite{rosettafold,rfdiffusion,proteinmpnn}, which proves to be effective tools to assist the development of small and large molecule drugs. Peptide drugs occupy a unique chemical and pharmacological space between small and large molecules, but the AI-assisted peptide drug discovery methods remain limited compared with those established for small molecules and proteins. Unbound peptide chains are usually at high free energy and entropy values, thus showing unstable conformations, while they trigger pharmacological effects when binding to specific receptors, forming a complex with equilibrium structures that are composed of a pair of receptor and ligand. \cite{marullo2013peptide, seebach2006helices}. Therefore, we focus on the designation of peptide drugs that can bind to specific receptors \cite{Todaro2023TargetingPT}. 
Recently, structure-based drug design (SBDD) methods are developed for target-aware small molecule generation \cite{Peng2022Pocket2MolEM, targetdiff,Lin2022DiffBPGD}, while these methods cannot be simply transferred to peptide design tasks for the following reasons: (i) Peptide drugs are usually of larger molecular weights ($500-5000$ Da) compared with small molecule drugs ($<1000$ Da) \cite{Francoeur20203DCN}; (ii) Topologies of atom connectivity in peptides are close to proteins rather than molecules; (iii) Internal geometries in peptides are of different patterns from small molecules.

Hence, we identify four challenges for the target-aware peptide drug design. \textit{First}, how to extract sufficient information from the conditional receptor contexts. This is indispensable for the model to generalize to new receptors. \textit{Second}, how to generate valid peptides that satisfy physicochemical rules. If the generated peptides are not chemically valid, other drug properties are unnecessary for further consideration. \textit{Third}, searching for natural peptides and replacing them with animal homologs, such as the discovery of insulin, GLP-1 and somatostatin, were the important strategies used for peptide drug discovery \cite{Kelly2022EffectsOG}. Therefore, instead of de novo design, the model should apply to another scenario: optimizing natural peptides to drugs with higher binding affinities. \textit{Besides}, there are few available benchmark datasets large enough to support the training of deep learning models for structure-based peptide drug design tasks, and it is urged to collect and organize a high-quality dataset that satisfies the demand for massive data.

\emph{(i) To address data shortage}, we first construct a dataset consisting of a large number of protein-peptide complexes through a series of systematic steps. \emph{(ii) To give a solution to the task}, we propose a generative model based on flow matching, to learn the structure distribution of peptides from their torsion geometry and the distributions of global translation and orientation. Instead of directly learning the explicit probability, \textsc{PPFlow} fits the gradient fields of the variables' evolutionary process from a prior distribution to the peptides' sequence-structure distributions and uses the learned gradient field to iteratively push the sequences and structures toward data distribution by an ordinary differential equation. \emph{(iii) To sufficiently extract representation from the protein}, we employ neural networks as the approximation to the gradient fields, parameterized with attention-based architectures to encode contextual proteins' structure and sequence. \emph{(iv) To confirm the validity of the model}, extensive experiments on multi-tasks are conducted. \textsc{PPFlow} reaches state-of-the-art performance compared with diffusion-based methods we transferred in generating peptides with higher affinity, stability, validity, and novelty. In optimization tasks, \textsc{PPFlow} also shows great superiority. Besides, experiments on flexible re-docking and side-chain packing demonstrate the models' potential in structure modeling. 

To sum up, our contributions include: (i) \textbf{New Task}. We consider the conditional generation of peptide drugs targeted at protein receptors as contexts, which is an area that has rarely been explored by deep learning. Other tasks including peptide optimization, flexible re-docking, and side-chain packing are also studied and evaluated. (ii) \textbf{Novel Model}. To fulfill the task, we establish a new model called \textsc{PPFlow} as a complete solution to model the internal geometries of torsion angles in peptides, achieving competitive performance in the discussed tasks. In addition, to our best knowledge, we are the first to establish flow-matching models for torsional angles lying on torus manifolds and to propose a deep-learning model for target-specific peptide generation. (iii) \textbf{Dataset Establishment}. For training the deep learning model, we collect a benchmark dataset called \texttt{PPBench2024} consisting of 15593 protein-peptide pairs with high quality.  Compared with previous protein-peptide datasets \cite{Agrawal2019BenchmarkingOD, Martins2023PropediaVA,Wen2018PepBDBAC}, \texttt{PPBench2024} augments and expands available data, and filters large amounts of low-quality data according to strict criteria, thus allowing deep-learning-based models to be trained on the datasets and to fulfill the peptide drug discovery tasks.

\section{Background}
\subsection{Problem Statement}
For a binding system composed of a protein-ligand pair (\emph{i.e.} protein-peptide pair) as $\mathcal{C}$, which contains $N_{\mathrm{pt}}$ amino acids of the protein and $N_{\mathrm{pp}}$ amino acids of the peptide, we represent the index set of the amino acids in the peptide as $\mathcal{I}_\mathrm{pp} =\{1,\ldots,N_{\mathrm{pp}}\}$, and the protein's amino acids as $\mathcal{I}_\mathrm{pt} = \{N_{\mathrm{pp}}+1,\ldots,N_{\mathrm{pp}} +N_{\mathrm{pt}} \}$. The amino acids can be represented by its type $s^{(i)}$, atom coordinates $X^{(i)} = (\bm{\mathrm{x}}^{(i,1)},\ldots, \bm{\mathrm{x}}^{(i,M^{(i)}})$, where $s^{(i)} \in \{1, \ldots, 20\}$, $\bm{\mathrm{x}}^{(i,j)} \in \mathbb{R}^{3}$, and $M^{(i)} = M(s^{(i)})$, meaning that the number of atoms is decided by the amino acid types. The sets of $\{s^{(i)}\}_{i=1}^{N_{\mathrm{pp}}}$ refers to a peptide sequence, and $\{X^{(i)}\}_{i=1}^{N_{\mathrm{pp}}}$ refers to its structure, which is the same for protein.

Therefore, $\mathcal{C} = \{(s^{(i)}, X^{(i)})\}_{i=1}^{N_\mathrm{pt} + N_{\mathrm{pp}}}$ can be split into two sets as $\mathcal{C} =\mathcal{P} \cup \mathcal{L}$, where $\mathcal{P} = \{(s^{(i)}, X^{(i)}): i\in \mathcal{I}_\mathrm{pt}\}$ and $\mathcal{L} = \{(s^{(i)}, X^{(i)}): i\in \mathcal{I}_\mathrm{pp}\}$. For \textit{protein-specific peptide generation}, our goal is to establish a probabilistic model to learn the distribution of molecules conditioned on the target proteins, \textit{i.e.} $p(\mathcal{L}|\mathcal{P})$.

\subsection{Riemanian Flow Matching}
Let $\mathcal{P} = \mathcal{P}(\mathcal{M})$ be the probability function space over a manifold $\mathcal{M}$ with Riemanian metric $g$. $q$ is probability distribution of data $x\in \mathcal{M}$, and $p$ is the prior distribution. The probability  path on $\mathcal{M}$ as  an interpolation in probability space written as $p_t: [0,1] \rightarrow \mathcal{P}$ satisfies $p_0 = p$ and $p_1 = q$. $u_t(x) \in \mathcal{T}_{x}\mathcal{M}$ is the corresponding gradient vector of the path on $x$ at time $t$. A Flow Matching (FM)  tangent vector field $v_t: [0,1] \times \mathcal{M} \rightarrow  \mathcal{M}$ is used to approximate $u_t(x)$ with the objective $L_{\mathrm{RFM}}(\theta) = \mathbb{E}_{t, p_t(x)} \|v_t(x) - u_t(x)\|^2_{g}$, in which $\theta$ are the parameters in $v_t$.
 $u_t$ is intractable, and an alternative is to construct conditional density path $p_t(x|x_1)$ whose gradient reads $u_t(x|x_1)$, and use Conditional Flow Matching (CFM) objective as   
\begin{equation}\vspace{-0.5em}
    L_{\mathrm{CRFM}}(\theta) = \mathbb{E}_{t\sim\mathcal{U}(0,1), p_1(x_1), p_t(x|x_1)} \|v_t(x) - u_t(x|x_1)\|^2_{g}, \notag
\end{equation}
since FM and CFM objectives have the same gradients as shown in \cite{fm, rfm}. Once the gradient field $v$ is learned, one can use ordinary differential equations $\frac{d}{dt}\varphi_t(x) = v_t(\varphi_t(x))$ with $\varphi_0(x) = x \sim p_0$ and $\varphi_t(x) = x_t$ to push the random variable $x_t \in \mathcal{M}$ from prior distribution $p_0$ to the data distribution $p_1$.
\section{Proposed Method}
\subsection{Peptide Structure Parameterization}

\begin{wrapfigure}{r}{0.5\linewidth}\vspace{-1.5em}
   \begin{center}
     \includegraphics[width=1.1\linewidth, trim = 00 00 00 01,clip]{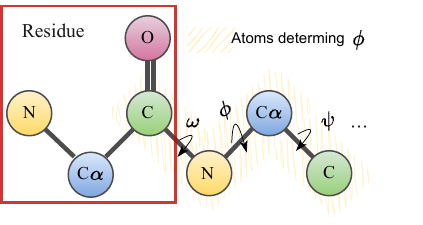}
   \end{center}\vspace{-1.9em}
   \caption{Backbone atom parameterization.} \label{fig:peptidebb}\vspace{-1.2em}
 \end{wrapfigure}

\label{sec:parametrize}
There are several parameterizations for peptide structures. Here, we firstly focus on the four backbone atoms, \emph{i.e.} $\{$N, C$\alpha$, C, O$\}$. One effective parameterization follows \textsc{AlphaFold2} \cite{alphafold}, in which the four atom positions $X^{(i)*} = \{\bm{\mathrm{x}}^{(i,1)}, \ldots, \bm{\mathrm{x}}^{(i,4)}\}$ form a rigid body, and represent $X^{(i)*}  \triangleq (\bm{\mathrm{x}}^{(i,2)}, O^{(i)})$ as variables on $\mathrm{SE}(3)$, in which $\bm{\mathrm{x}}^{(i,2)} \in \mathbb{R}^3$ as C$\alpha$'s coordinate is the translation vector and $O_{i} \in \mathrm{SO}(3)$ is an rotation matrix obtained by the ideal frame constructed by $X^{(i)*}$, leading $\{(\bm{\mathrm{x}}^{(i,2)}, O^{(i)})\}_{i=1}^{N_{\mathrm{pp}}}$ to represent the backbone structure. Hence, the structure representation lies on a manifold with $6N_{\mathrm{pp}}$-degree of freedom.   However, while the intra-residues' internal geometries are fixed, \emph{e.g.} bond lengths of `N-C$\alpha$' and `C$\alpha$-C', and bond angles of `N-C$\alpha$-C', the other inflexible inter-residues' geometries of bond lengths such as `C-N' and angles like `C-N-C$\alpha$' are not constrained in the parameterization. 

In comparison, we focus on the torsion angles of $\phi$, $\psi$, and $\omega$ (shown in Figure.~\ref{fig:peptidebb}) as redundant geometries according to the physicochemical conclusions \cite{Padmanabhan2014HandbookOP} and our observations (See Appendix.~\ref{app:geomanaly}). While the local structures can be reconstructed with torsion angles and ideal bond lengths and angles with \textsc{NeRF} \cite{parson2005jcc}, the global translation and orientation have to be determined to obtain the peptides' poses relative to the target proteins \cite{corso2023diffdock}. By this means, the backbone peptide structure can be represented as $\{X^{(i)*}\}_{i=1}^{N_{\mathrm{pp}}} \triangleq\{(\phi^{(i)}, \psi^{(i)}, \omega^{(i)})\}_{i=2}^{N_\mathrm{pp}} \cup \{\bm{\mathrm{x}}^{(\mathrm{C})}, O^{(\mathrm{C})}\}$. Note that $i$ start from $2$ because four consecutive atoms form a torsion angle,  $\bm{\mathrm{x}}^{(\mathrm{C})}$ is the global translation of the peptide \underline{c}entroid, and $O^{(\mathrm{C})}$ is the rotation matrix of the global frame constructed by $\{\bm{\mathrm{x}}^{(1,1)}, \bm{\mathrm{x}}^{(\mathrm{C})}, \bm{\mathrm{x}}^{(N_{\mathrm{pp}}, 3)}\}$. 
Hence, the backbone structure is parameterized by variables on $\mathbb{T}^{3N_{\mathrm{pp}}-3} \times \mathrm{SE}(3)$. The advantages of the parametrization include \emph{(i)} the inflexible bond lengths and angles are set to be constants, avoiding broken or overlapped bonds as unrealistic structures and \emph{(ii)} fewer degrees of freedom to reduce the task difficulties ($3N_{\mathrm{bb}} + 3 < 6N_{\mathrm{bb}}$). 

In the following, the CFM on the discussed manifolds will be described in Sec.~\ref{sec:torusflow} and Sec.~\ref{sec:se3flow}, and for the sequence type CFM, it will be discussed in Sec.~\ref{sec:typeflow}.
Besides, for the other atoms in side-chains, we use the rotamer $\chi$-angles as the parameterization, discussed in Sec.~\ref{sec:rotamer}.

\subsection{\textsc{PPFlow} on Torus}
\begin{figure*}[ht]
    \centering
    \includegraphics[width=0.95\linewidth, trim = 150 120 470 100,clip]{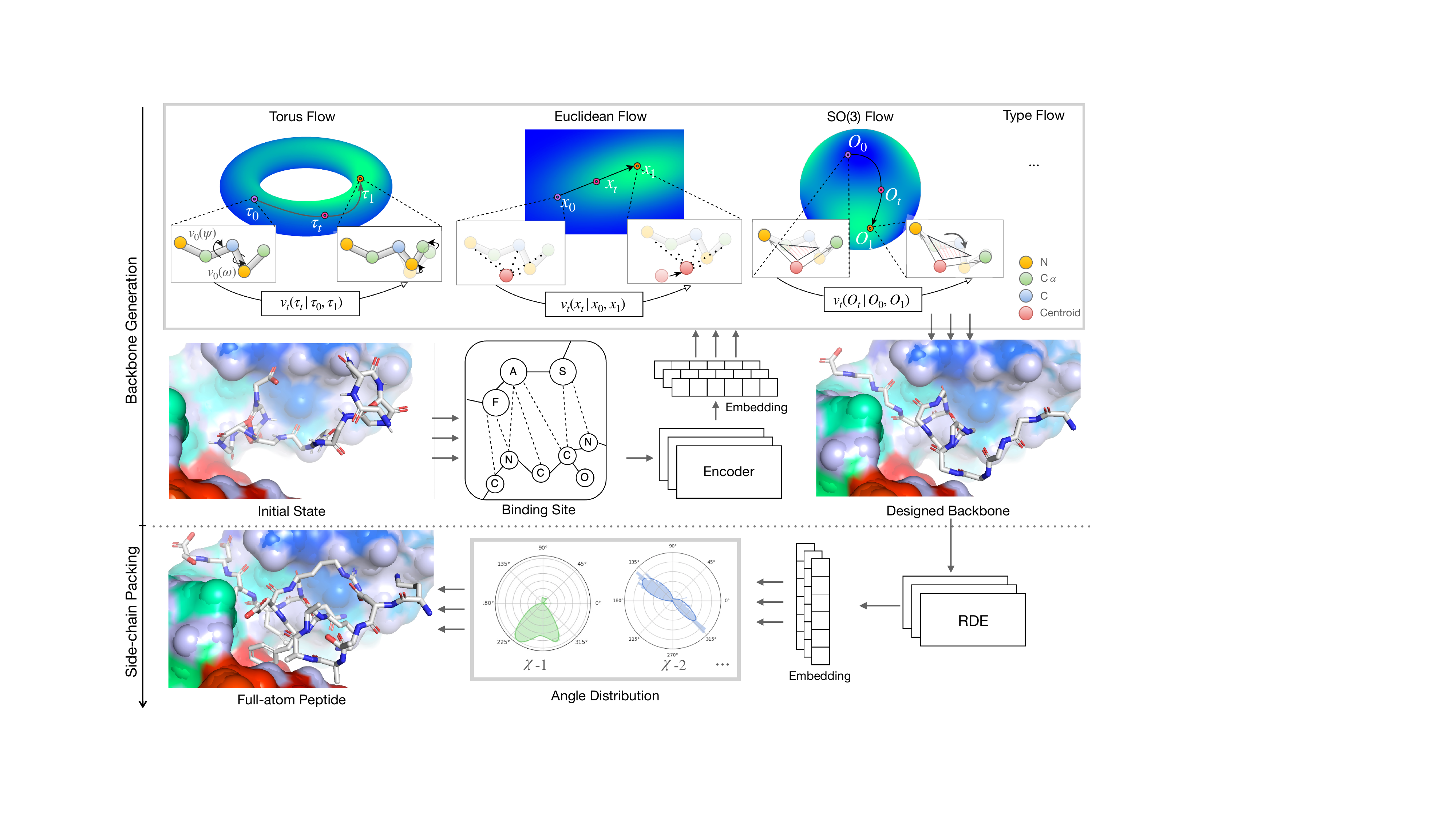}
        \caption{Workflows of \textsc{PPFlow} in target-aware peptide drug generation task.} \label{fig:workflow}
\end{figure*}
\label{sec:torusflow}
For each torsion angle lies in $[-\pi, \pi)$,  $N$ torsion angles of a structure define a hypertorus $\mathbb{T}^N$, with $N=3N_{\mathrm{pp}} - 3$. To construct a conditional flow on $\mathbb{T}^N$, we follow RCFM, to obtain $\bm{\tau}_t$ using the geodesic connecting $\bm{\tau}_0$ and $\bm{\tau}_1$ with
\begin{equation} 
\bm{\tau}_t = \exp_{\bm{\tau}_0}(t\log_{\bm{\tau}_0}(\bm{\tau}_1)),    
\end{equation}
where $\bm{\tau}_0 \sim p_0$, $\bm{\tau}_1 \sim p_1$ and $\bm{\tau}_t \in \mathbb{T}^N$, and the conditional vector field $u_t(\bm{\tau}|\bm{\tau}_1) = \frac{d}{dt}\bm{\tau}_t= \bm{\dot \tau}_t$. $\log_{\bm{\tau}_0}: \mathbb{T}^N \rightarrow \mathcal{T}_{\bm{\tau}_0}\mathbb{T}^N$ is the logarithm map, projecting points on $\mathbb{T}^N$ to the tangent space of $\bm{\tau}_0$; $\exp_{\bm{\tau}_0}:  \mathcal{T}_{\bm{\tau}_0}\mathbb{T}^N \rightarrow \mathbb{T}^N $ is the exponential map, projecting points back to torus \cite{rfm}. 

For the torus, we parameterize the manifold as the quotient space $\mathbb{R}^{N} / 2\pi \mathbb{Z}^N$, leading to the equivalence relations $\bm{\tau} = (\tau^{(1)}, \ldots, \tau^{(N)}) \cong (\tau^{(1)} + 2\pi, \ldots, \tau^{(N)}) \cong (\tau^{(1)}, \ldots, \tau^{(N)} + 2\pi)$ \cite{jing2023torsional}. By this means the prior distribution $p_0$ is chosen as a product of standard wrapped normal distribution on $\mathbb{R}^N$ as 
\begin{equation}
\begin{aligned}
    p_0 (\bm{\tau}) =& \mathcal{WN}(\bm{\tau}) \propto \prod_{i=1}^N\sum_{d \in \mathbb{Z}} \exp \left( -\frac{\|{\tau}^{(i)} + 2\pi d\|^2}{2}\right).
\end{aligned}
\end{equation}
To construct the conditional path, considering the flow 
\begin{equation}
\bm{\tau}_t = \sigma_t(\bm{\tau}_1) \bm{\epsilon} + \bm{\mu}_t(\bm{\tau}_1),  \label{eq:wnpath}
\end{equation}
where $\bm{\epsilon} \sim \mathcal{WN}(\bm{\epsilon})$. To build the gradient vector, we propose that 
\textbf{Theorem~3.} in \cite{fm} still holds for the defined path, as the following proposition:
\begin{proposition} \label{prop1}
Let $p_t(\bm{\tau}|\bm{\tau}_1)$ be probability path as in Equation.~\ref{eq:wnpath}. Its vector field has the form
\begin{equation}
    u_t(\bm{\tau}|\bm{\tau}_1) = \frac{\dot\sigma_t(\bm{\tau}_1)}{\sigma_t(\bm{\tau}_1)} \left(\bm{\tau} - \bm{\mu}_t(\bm{\tau}_1)\right) +  \bm{\dot\mu}_t(\bm{\tau}_1).
\end{equation}
Hence, $u_t(\bm{\tau}|\bm{\tau}_1)$ generates the probability path $p_t(\bm{\tau}|\bm{\tau}_1)$.
\end{proposition}
The proof is given in Appendix.~\ref{app:proof1}. 
Besides, the following proposition ensures the correctness of the learning objective in Conditional Torus Flow Matching with proof in Appendix.~\ref{app:proof2}:
\begin{proposition} \label{prop2}
    Given $p_t(\bm{\tau})$, $ \bm{\tau} \in \mathbb{T}^N$, the conditional and unconditional flow matching losses have equal gradients \emph{w.r.t.} $\theta$: $\nabla_{\theta}L_{\mathrm{UTFM}}(\theta) = \nabla_{\theta}L_{\mathrm{CTFM}}(\theta)$.
\end{proposition}

Following \cite{otfm}, we in practice construct our flow-matching objective in an Independent Coupling way:
\begin{equation}
\begin{aligned}
 \bm{\tau}_t(\bm{\tau}_0, \bm{\tau}_1) &= \exp_{\bm{\tau}_0}(t\log_{\bm{\tau}_0}(\bm{\tau}_1)) + \sigma\bm{\epsilon};\\
 u_t(\bm{\tau}|\bm{\tau}_0, \bm{\tau}_1) &= \frac{d}{dt} \exp_{\bm{\tau}_0}(t\log_{\bm{\tau}_0}(\bm{\tau}_1)),
\end{aligned}
\end{equation}
which is a specific instance of the above proposition. The explicit equations for geodesics on a hypertorus $\mathbb{T}^N$ can be complex and depend on the specific parametrization chosen for the hypertorus \cite{jantzen2012geodesics}. Instead, we regard the torsion angles as mutually orthogonal and their interpolation paths on the torus are linear \emph{w.r.t.} $t$ in each direction. Therefore, we employ a fast implementation of calculating $\bm{\mu}_t \approx \exp_{\bm{\tau}_0}(t\log_{\bm{\tau}_0}(\bm{\tau}_1))$ by
\begin{equation}
\begin{aligned}
    \bm{\mu}_t (\bm{\tau}|\bm{\tau}_0, \bm{\tau}_1) &= t \bm{\tau}'_1 + (1-t)\bm{\tau}_0; \\
    u_t(\bm{\tau}|\bm{\tau}_0, \bm{\tau}_1)  &= \bm{\tau}'_1 - \bm{\tau}_0,
\end{aligned}
\end{equation}
where $\bm{\tau} \cong \bm{\tau}'_1 = (\bm{\tau}_1 - \bm{\tau}_0 + \pi)\bmod (2\pi) - \pi$. The geodesic distance is approximated with the euclidean ones, as $\|\bm{\tau}_1 - \bm{\tau}_0\|_{\mathbb{T}^N}^2 \approx\|\bm{\tau}'_1 - \bm{\tau}_0\|_2^2$, which satisfies definitions of premetric in Sec.~3.2 in \cite{rfm}. This leads the closed-from expression of the loss to train the conditional Torus Flow Matching to \vspace{-0.5em}
\definecolor{mygray}{gray}{0.95}
\begin{center}\vspace{-0.5em}			
    \colorbox{mygray} {		
      \begin{minipage}{0.977\linewidth} 	
       \centering
       \vspace{-0.3em}
\begin{equation}\label{eq:losstfm}
    \small
    L_{\mathrm{TFM}}(\theta) = \mathbb{E}_{\substack{t\sim\mathcal{U}(0,1), p_1(\bm{\tau}_1),\\ p_0(\bm{\tau}_0), p_t(\bm{\tau}|\bm{\tau}_0, \bm{\tau}_1)}} \|v_t(\bm{\tau}) - \bm{\tau}'_1 + \bm{\tau}_0\|^2_2.
\end{equation}      
      \end{minipage}}			
      \vspace{-1em}
\end{center}
\normalsize

\subsection{\textsc{PPFlow} on SE(3)}
\label{sec:se3flow}
The torsion angles can reconstruct the local structures of the peptides, while in the global coordinate system, to represent the positions of residues, we need to determine their global translations and rotations.

As discussed in Sec.~\ref{sec:parametrize}, the pose representation on $\mathrm{SE}(3)$ can be decomposed into global translation as $\bm{\mathrm{x}}^{(\mathrm{C})} \in \mathbb{R}^3$ and rotation $O^{(\mathrm{C})} \in \mathrm{SO}(3)$. For notation simplicity, we omit the superscript $^{(\mathrm{C})}$ in this part. To model the probability path of $\bm{\mathrm{x}}_t$, we employ vanilla Gaussian CFM on Euclidean manifolds, with Independent Coupling techniques:
\begin{equation}\vspace{-0.5em}
\begin{aligned}
        \bm{\mu}_t (\bm{\mathrm{x}}|\bm{\mathrm{x}}_0,\bm{\mathrm{x}}_1) &=  t\bm{\mathrm{x}}_1 + (1-t)\bm{\mathrm{x}}_0;\\
        \sigma_t &= \sigma,     
\end{aligned}
\end{equation}
where $\bm{\mathrm{x}}_0 \sim \mathcal{N}(\bm{\mathrm{x}}| \bm{0}, I)$, thus leading to the Gaussian probability path of $p_t(\bm{\mathrm{x}}|\bm{\mathrm{x}}_0,\bm{\mathrm{x}}_1)=\mathcal{N}(\bm{\mathrm{x}}|\bm{\mu}_t, \sigma_t)$, and the loss to train the conditional Euclidean Flow Matching as
\vspace{-1em}
\definecolor{mygray}{gray}{0.95}
\begin{center}\vspace{-0.em}			
    \colorbox{mygray} {		
      \begin{minipage}{0.977\linewidth} 	
       \centering
        \vspace{-1em}
\begin{equation}\label{eq:lossefm}
    \small
    L_{\mathrm{EFM}}(\theta) = \mathbb{E}_{\substack{t\sim\mathcal{U}(0,1), p_1(\bm{\mathrm{x}}_1), \\p_0(\bm{\mathrm{x}}_0), p_t(\bm{\mathrm{x}}|\bm{\mathrm{x}}_0, \bm{\mathrm{x}}_1)}} \|v_t(\bm{\mathrm{x}}) - \bm{\mathrm{x}}_1 + \bm{\mathrm{x}}_0\|^2_2.
\end{equation}      
      \end{minipage}}			
      \vspace{-1em}
\end{center}
To model the probability path of rotation matrix $O_t\in\mathrm{SO(3)}$, we employ  $\mathrm{SO}(3)$-CFM \cite{yim2023fast}, in which $O_t = \exp_{O_0}(t\log_{O_0}(O_1))$. In the implementation, since the  $\mathrm{SO}(3)$ is a simple manifold with closed-form geodesics, the exponential map can be computed using Rodrigues’ formula and the logarithmic map is similarly easy to compute with its Lie algebra $\mathfrak{so}(3)$ \cite{yim2023se3}. The prior distribution to sample $O_0$ is defined as isotropic Gaussian distribution, by first parameterizing $O_0$ in axis-angle, where the axis of rotation is sampled uniformly and the density function of rotation angle $\vartheta$ reads $\mathcal{IG}(\vartheta) = \frac{1-cos\vartheta}{\pi}\sum_{l=0}^\infty(2l+1)e^{-l(l+1)\epsilon}\frac{\sin((l+\frac{1}{2})\vartheta)}{\sin(\vartheta/2)}$ \cite{Creasey2017FastGO}.
For the conditional gradient field, we employ fast numerical tricks \cite{Bose2023SE3StochasticFM} to calculate the  $\mathrm{SO}(3)$ component of the global Orientation conditional Flow Matching objectives:
\vspace{-1em}
\definecolor{mygray}{gray}{0.95}
\begin{center}\vspace{-0.5em}			
    \colorbox{mygray} {		
      \begin{minipage}{0.977\linewidth} 	
       \centering
        \vspace{-1em}
\begin{equation}\label{eq:lossofm}
    \small
    L_{\mathrm{OFM}}(\theta) = \mathbb{E}_{\substack{t\sim\mathcal{U}(0,1), p_1(O_1), \\p_0(O_0), p_t(O|O_0, O_1)}} \left\|v_t(O) - \frac{\log_{O_t}(O_0)}{t}\right\|^2_{\mathrm{SO(3)}}.
\end{equation}      
      \end{minipage}}			
      \vspace{-1em}
\end{center}
The correctness of the learning objective of Equation.~\ref{eq:lossefm} and ~\ref{eq:lossofm} is proposed and proven in \cite{fm} and \citep{Bose2023SE3StochasticFM}, respectively.

\subsection{\textsc{PPFlow} on Amino Acid Types}
\label{sec:typeflow}
For the amino acid sequence $\bm{s} = \{s^{(i)}\}_{i=1}^{N_\mathrm{PP}}$, we directly model the probability vector of each type, where $\bm{c}^{(i)}_t$ is defined as the probability vector of multinomial distribution with $s^{(i)}_t \sim P(\bm{c}^{(i)}_t)$. We omit $^{(i)}$ for notation simplicity. To build a path, we define $\bm{c}_1 = \mathrm{onehot}(s_i)$, and $\bm{c}_0 = (\frac{1}{20}, \ldots, \frac{1}{20})$. Analogously to the Euclidean flow, we define  $\bm{c}_t = t\bm{c}_1 + (1-t)\bm{c}_0$, and $u_t(\bm{c}|\bm{c}_0, \bm{c}_1) = \bm{c}_1 - \bm{c}_0$.
It is easy to prove $\bm{c}_t$ is a probability vector since its summation equals 1 across all types. Instead of an MSE training loss in $\mathbb{R}^{20}$, we, inspired by multinomial diffusion \cite{hoogeboom2021argmax}, propose a multinomial flow matching objective:
\vspace{-1em}
\definecolor{mygray}{gray}{0.95}
\begin{center}\vspace{-1.0em}			
    \colorbox{mygray} {		
      \begin{minipage}{0.977\linewidth} 	
       \centering
       \vspace{-1em}
\begin{equation}
    \small L^{(i)}_{\mathrm{SFM}}(\theta) = \mathbb{E}_{\substack{t\sim\mathcal{U}(0,1), p_1(\bm{c}_1), \\p_0(\bm{c}_0), p_t(\bm{c}|\bm{c}_0, \bm{c}_1)}} \mathrm{CE}\left(\bm{c} + (1-t)v_t(\bm{c}), \bm{c}_1\right), \notag
\end{equation}      
      \end{minipage}}			
      \vspace{-1em}
\end{center}
where $\mathrm{CE}(\cdot,\cdot)$ is cross-entropy, and the loss directly measures the difference between the true probability and the inferred one $ \bm{\hat c}_1 = \bm{c} + (1-t)v_t(\bm{c})$.
For the whole sequence, the loss function is a summation of $L^{(i)}_{\mathrm{SFM}}(\theta)$ over $i$, as 
\begin{equation}\vspace{-1em}
 L_{\mathrm{SFM}}(\theta) = \sum_{i=1}^{N_\mathrm{pp}}L^{(i)}_{\mathrm{SFM}}(\theta).\label{eq:losssfm}
\end{equation}

\subsection{Overall Training Loss}
The overall loss function is the summation of the four loss functions in Equation.~\ref{eq:losstfm}, \ref{eq:lossefm}, \ref{eq:lossofm} and \ref{eq:losssfm}. To fully utilize the sufficient contextual information as conditional inputs, for each flow-matching vector field $v_t(\cdot)$, we instead use all the structure-sequence contexts as input. In this way, we will first sample $\bm{\tau}_t$, $\bm{\mathrm{x}}_t^{(\mathrm{C})}$, $O_t^{(\mathrm{C})}$ and $\bm{s} = \{s^{(i)}_t\}_{i=1}^{N_\mathrm{pp}}$ through the defined conditional probability paths, and then employ \textsc{NeRF} \cite{parson2005jcc} to reconstruct the sequence and backbone structure in local frame, and translate and rotate the structures such that its centroid coordinate and orientation equals to  $\bm{\mathrm{x}}_t^{(\mathrm{C})}$ and $O_t^{(\mathrm{C})}$. Thus, we can obtain $\mathcal{L}^*_t = \{(s_t^{(i)}, X_t^{(i)*}): i\in \mathcal{I}_\mathrm{pp}\} $. Finally, the whole binding complex at $t$ is used as inputs of $v_t(\cdot)$, as $v_t(\cdot| \mathcal{L}^*_t \cup \mathcal{R}) = v_t(\cdot| \mathcal{C}^*_t) $. Therefore, our overall training loss is written as 
\vspace{-1.em}
\definecolor{mygray}{gray}{0.95}
\begin{center}\vspace{-0.5em}			
    \colorbox{mygray} {		
      \begin{minipage}{0.977\linewidth} 	
       \centering
        \vspace{-1em}
\begin{equation}\label{eq:lossafm}
    \small
    L_{\mathrm{PPF}} = \mathbb{E}_{\mathcal{C}^*} (L_{\mathrm{TFM}} +L_{\mathrm{EFM}}+ L_{\mathrm{OFM}} + L_{\mathrm{SFM}}),
\end{equation}      
      \end{minipage}}			
      \vspace{-1em}
\end{center}
\normalsize
in which $L_{\mathrm{TFM}} = \mathbb{E}_{\substack{t\sim\mathcal{U}(0,1), p_1(\bm{\tau}_1),\\ p_0(\bm{\tau}_0), p_t(\bm{\tau}|\bm{\tau}_0, \bm{\tau}_1)}} \|v_t(\bm{\tau}; \mathcal{C}^*) - \bm{\tau}'_1 + \bm{\tau}_0\|^2_2$, and the other three can be written in a similar way. We write $\mathcal{C}^* = \mathrm{rec}(\bm{\tau}, \bm{\mathrm{x}}^{(\mathrm{C})},O^{(\mathrm{C})}, \bm{s} )$, meaning the input binding complexes are reconstructed from the four variables.

\subsection{Sampling with ODE}
\label{sec:sampling}
For the trained flow-matching vector fields, we sampling process is the solution of the following ordinary
the differential equation as $\frac{d}{dt}\mathcal{C}^* = v_t(\mathcal{C}^*)$, in which the defined method is employed as the numerical solution:
\definecolor{mygray}{gray}{0.95}
\begin{center}\vspace{-1em}			
    \colorbox{mygray} {		
      \begin{minipage}{0.977\linewidth} 	
       \centering
       \vspace{-1.3em}
\begin{flalign}
\text{Step 1:}   \quad\quad\bm{\tau}_{t+\Delta t} &= \mathrm{reg}(\bm{\tau}_t + v_t(\bm{\tau}_t; \mathcal{C}^*_t)\Delta t);&\notag\\
\bm{\mathrm{x}}^{(C)}_{t+\Delta t} &= \bm{\mathrm{x}}_t + v_t(\bm{\mathrm{x}}_t^{(C)}; \mathcal{C}^*_t\Delta t);&\\
O^{(\mathrm{C})}_{t+\Delta t} &= O^{(\mathrm{C})}_t\exp_{O_t^{(\mathrm{C})}}\left( v_t(O^{(\mathrm{C})}_t; \mathcal{C}^*_t)\Delta t\right);&\notag\\
\bm{c}^{(i)}_{t+\Delta t}&= \mathrm{norm}(\bm{c}^{(i)}_t + v_t(\bm{c}^{(i)}_t; \mathcal{C}^*_t)\Delta t)  & \notag\\
s^{(i)}_t &\sim \bm{c}^{(i)}_t \quad \quad\quad\quad 1\leq i\leq N_\mathrm{pp};&  \notag
\end{flalign}
      \end{minipage}}			
      \vspace{-1em}
\end{center}

\definecolor{mygray}{gray}{0.95}
\begin{center}\vspace{-1em}			
    \colorbox{mygray} {		
      \begin{minipage}{0.977\linewidth} 	
       \centering
       \vspace{-1.3em}
       
\begin{flalign}
&\small\text{Step 2:} \quad \mathcal{C}^*_{t+\Delta t} = \mathrm{rec}(\bm{\tau}_{t+\Delta t}, \bm{\mathrm{x}}^{(\mathrm{C})}_{t+\Delta t},O_{t+\Delta t}^{(\mathrm{C})}, \bm{s}_{t+\Delta t}),&
\end{flalign}
      \end{minipage}}			
      \vspace{-1em}
\end{center}

where $\mathrm{norm}(\cdot)$ means normalizing the vector to a probability vector such that the summation is $1$, and $\mathrm{reg}(\cdot)$ means regularize the angles by $\mathrm{reg}(\bm{\tau}) =(\bm{\tau}+\pi)\bmod{(2\pi)} - \pi $ .
\subsection{Parameterization with Neural Networks}

\textbf{Encoder.} We first adopt two multi-layer perceptrons (MLPs). The embedding of single residues is obtained by one MLP, which encodes the residue type, backbone dihedral angles, and local atom coordinates. The other MLP for residue pairs encodes the distance, the dihedral angles between the $i$ and $j$ residues, and their relative positions. Then, we stack 6-layer transformers with self-attention mechanisms used to update the inter-residues' embedding and obtain residues' embedding for each amino acid by $\bm{h}_i \in \mathbb{R}^{D}$. 

 \textbf{Equivariance.} The conditional distribution of $p(\mathcal{L}|\mathcal{R})$ have to be roto-translational equivariant to ensure the generalization, \emph{i.e.} $ p(O\mathcal{L}+\bm{\mathrm{x}}|O\mathcal{R}+\bm{\mathrm{x}}) = p(\mathcal{L}|\mathcal{R})$, for $\bm{\mathrm{x}} \in \mathbb{R}^3$ and $O \in \mathrm{SO}(3)$, where $O\mathcal{L} + \bm{\mathrm{x}} = \{(s^{(i)}, OX^{(i)}+\bm{\mathrm{x}})\}$. To avoid the problem of non-existence of such a probability due to translation, we here adopt the zero-mass-center techniques \cite{Rudolph2020SameSB, yim2023se3}, by subtracting the mass center of the conditional receptor from all inputs' coordinates to the neural network \cite{targetdiff, Lin2022DiffBPGD, Satorras2021EnEN}, which also helps to improve the training stability. Further, in terms of CFM, the
following proposition indicates the roto-translational equivariance of each flow-matching vector field.
\begin{proposition} \label{prop3}
    Let $p_0(\bm{\tau})$, $p_0(\bm{\mathrm{x}}^{(\mathrm{C})})$, $p_0(O^{(\mathrm{C})})$, and $p_t(\bm{c})$ be SE(3)-invariant distribution, and the flow-matching vector field $v_t(\bm{\mathrm{x}}^{(\mathrm{C})}|\mathcal{C}^*)$ be SE(3)-equivariant, $v_t(O^{(\mathrm{C})}|\mathcal{C}^*)$ be T(3)-invariant and SO(3)-equivariant, and $v_t(\bm{c}|\mathcal{C}^*)$ and $v_t(\bm{\tau}|\mathcal{C}^*)$ be SE(3)-invariant, then the density $p(\mathcal{L}^*|\mathcal{R})$ generated by the ODE sampling process SE(3)-equivariant.
\end{proposition}

We employ \textsc{ITA} \cite{alphafold} with \textsc{LoCS} \cite{kofinas2022rototranslated} (For details see Appendix.~\ref{app:locs} ), to ensure the equivariance and the invariance of the vector fields in \ref{prop3}.

\subsection{Side-Chain Packing}\label{sec:rotamer}
For the complete solution to full-atom design, after the backbone atoms are generated, we employ a Rotamer Density Estimator (\textsc{RDE}) \cite{Luo2023RotamerDE} as the probabilistic model to model the side-chain rotamers $\{\bm{\chi}_i\}_{i=1}^{N_\mathrm{pp}}$.  It uses a Conditional Flow on $\mathbb{T}^{N_\mathrm{rt}}$ based on the rational quadratic spline flow \cite{durkan2019neural, rezende2020normalizing}, where $N_\mathrm{rt}$ is the total rotamer number. In practice, we use the pre-trained version of \textsc{RDE} capable of perceiving the side-chain conformations by training the model on large datasets of \texttt{PDB-REDO} \cite{Joosten2014ThePS}. Further, we conduct fine-tuning on our protein-peptide complex datasets. The effectiveness of the side-chain packing model named \textsc{RDE-PP}, transferred from protein-protein complexes to the protein-peptide ones, is empirically shown in Sec.~\ref{sec:exp:sidechain}.

The overall workflow of \textsc{PPFlow} including the backbone generation and side-chain packing is shown in Figure.~\ref{fig:workflow}.

\section{Related Work}
\textbf{Structure-based drug design.} Success in 3D molecule generation and increasing available structural data raises scientific interest in structure-based drug design (SBDD). Grid-based methods regard the Euclidean space as discrete and predict the molecules' structures on the grids \cite{luo20213d, masuda2020generating}. With great advances in Graph Neural Networks \cite{GNN1,GNN2,GNN3}, Equivariant neural networks have advanced the structure predictions and helped to improve the tasks greatly \cite{peng2022pocket2mol, liu2022graphbp}. Then, diffusion methods \cite{targetdiff, schneuing2022structurebased, Lin2022DiffBPGD} attempt to generate the ordered atoms' positions and types at full atom levels. Further, a series of works of fragment-based drug design are proposed, by mimicking classical drug design procedures to generate both atoms and functional groups that form molecule drugs \cite{flag,lin2023d3fg}. 

\textbf{Protein generation.} Machine learning methods for protein modeling have achieved great success in recent years \cite{protein1, protein2, protein3, Wu2023AHT}, and peptides can be regarded as fragments making up a protein. Techniques on protein backbone generation assist realistic peptide structure design. For example, FoldingDiff \cite{Wu2022ProteinSG} is also based on redundant angle generation, which employs diffusion models on torus to achieve backbone design. RFDiffusion and Chroma \cite{rfdiffusion,Ingraham2022IlluminatingPS} use different diffusion schemes, and achieve state-of-the-art generation performance on protein generations. Recently, protein backbone generation methods employ flow matching techniques, to explore the applicability and effectiveness \cite{yim2023fast, Bose2023SE3StochasticFM}. For protein side chains, methods usually focus on protein-protein complexes, such as RED-PPI \cite{Luo2023RotamerDE} and DiffPack \cite{ zhang2023diffpack}. Our side-chain packing methods follow RDE-PPI, and achieves a good generalization performance on peptides. 
\vspace{-0.0em}
\section{Experiment}
\subsection{Dataset}
\label{sec:data}

\vspace{-0.0em}
\begin{figure*}[t]
    \begin{minipage}{.70\textwidth}
\centering
\captionof{table}{Comparison for target-aware peptide generation. Values in \textbf{bold} are the best.} \label{tab:gencomp}\vspace{-0.8em}
\resizebox{1.0\columnwidth}{!}{
\begin{tabular}{lrrrrrr}
\toprule
          & $\Delta G (\downarrow)$  & IMP\%-B$(\uparrow)$ & IMP\%-S$(\uparrow)$ & Validity$(\uparrow)$ & Novelty$(\uparrow)$ & Diversity \\
          \midrule
\textsc{DiffPP}         & -319.54 & 28.75\%  & 3.72\%   & 0.41     & 0.89    & 0.28      \\
\textsc{DiffBP-PP}      & -221.06 & 19.34\%  & 1.04\%   & 0.42     & 0.76    & 0.41      \\
\midrule
\textsc{PPFlow-BB} & -349.59 & \textbf{36.02}\%  & 4.04\%  & \textbf{1.00}     & \textbf{0.99}    & \textbf{0.67}      \\
\textsc{PPFlow-FA}  & \textbf{-351.27}& 35.43\%  & \textbf{4.17\%}  & \textbf{1.00}     & \textbf{0.99}    & \textbf{0.67}     \\
\bottomrule
\end{tabular}
} 
\subfigure{ \label{fig:optbind}
    \includegraphics[width=0.47\linewidth, trim = 5 10 10 0,clip]{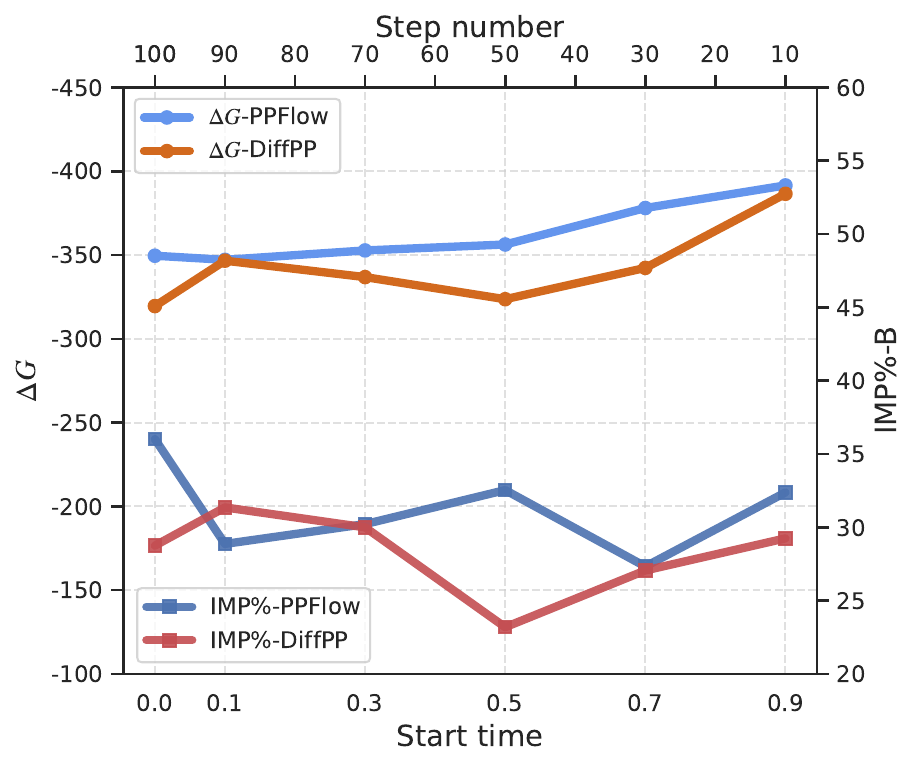}
    }\vspace{-0.5em}\hspace{-1mm}
    \subfigure{ \label{fig:optnovel} 
    \includegraphics[width=0.47\linewidth, trim = 5 10 10 00,clip]{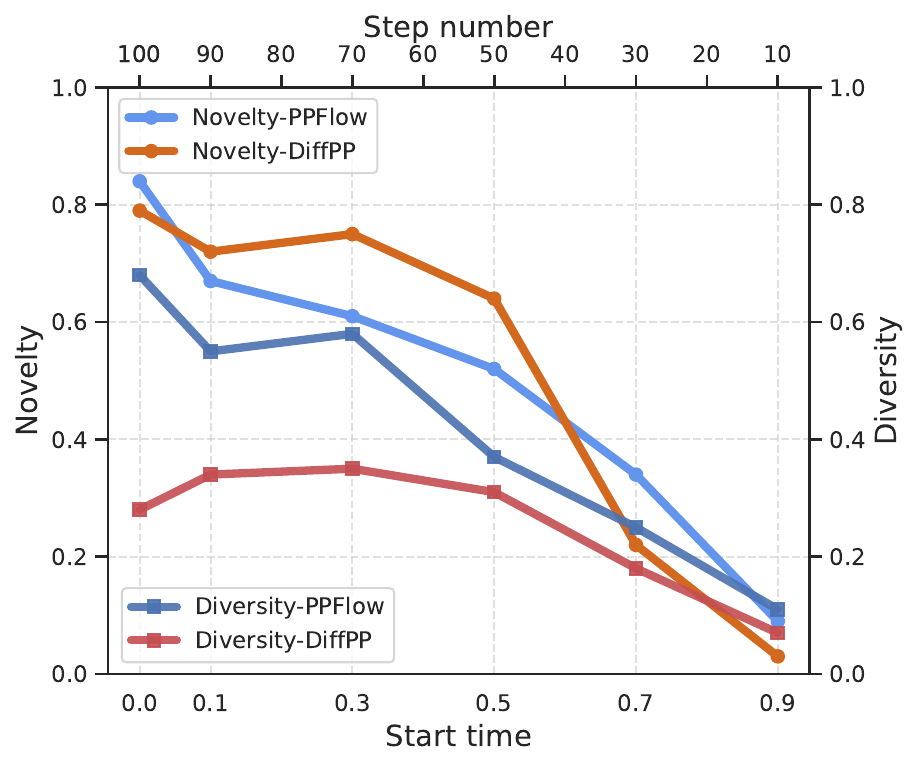}
    }\vspace{-0.5em}
    \captionof{figure}{Metrics for generated peptides of methods in different optimization steps.}\vspace{-1.5em}\hspace{2mm}\label{fig:optcomp}
\end{minipage}
    \begin{minipage}{.28\textwidth}\centering
    \includegraphics[width=0.9\linewidth, trim = 800 130 680 240,clip]{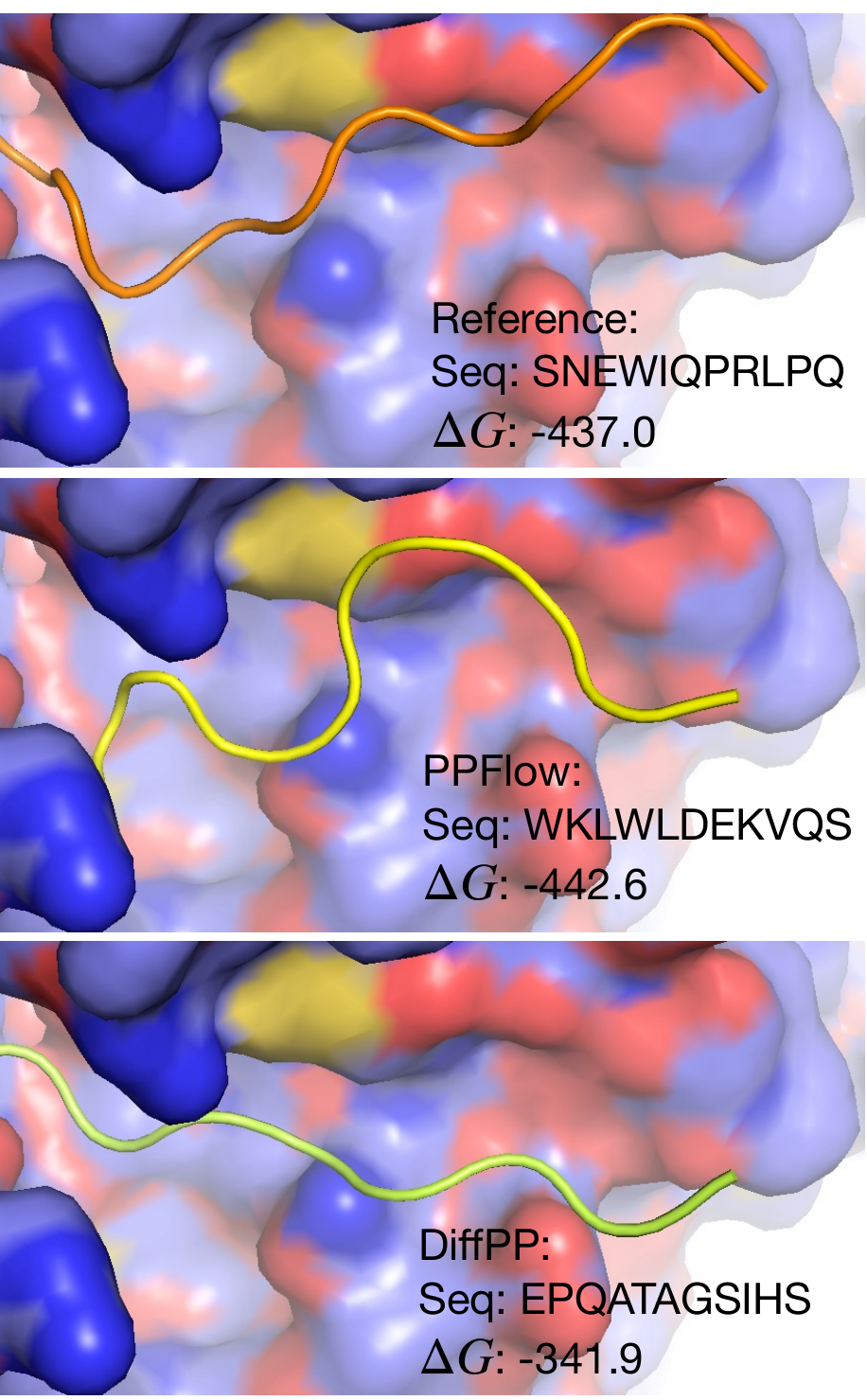}\vspace{-0.2em}
    \captionof{figure}{Peptides designed by different methods and reference.}\vspace{-1.5em} \label{fig:genvis}
    \end{minipage}\vspace{-0.2em}
\end{figure*}
\textbf{Training set.} To satisfy the need for massive data to train deep learning models, we construct \texttt{PPBench2024}, through a series of systematic steps: \textit{First}, we source complexes from the RCSB database \cite{Zardecki2016RCSBPD}, specifically selecting those containing more than two chains and excluding any with nucleic acid structures, and defining interactions between a pair as a minimum intermolecular distance of 5.0Å or less. \textit{Subsequently}, only those complexes featuring peptide chains that do not exceed 30 amino acids in length are included, to better mimic existing peptide drug sizes. \textit{Then}, the water molecules and heteroatoms are eliminated. \textit{Finally}, we select only peptide molecules composed entirely of amino acids, remove the modified peptides with functional groups other than amino acids, and filter peptides with broken bonds according to the ideal bond length, \emph{i.e.} the bond is unbroken if the observed bond length is between the ideal length plus or minus 0.5Å. The number of final screened complex instances of peptide-protein pairs is 9070. 
Further, we screen the existing datasets of \texttt{PropediaV2.3} \cite{Martins2023PropediaVA} and \texttt{PepBDB} \cite{Wen2018PepBDBAC} with the same criterion, leading to additional 6523 instances to expand it. Appendix.~\ref{app:dataprepro} gives details.
We split the \texttt{PPBench2024} into training and validation sets according to the clustering of the proteins that are closest to the peptide ligand via \textsc{MMSeqs2} \cite{steinegger2017mmseqs2} with the ratio of $9:1$.

\textbf{Test set.} To evaluate the model performance, we use an existing benchmark dataset called \texttt{PPDBench} \cite{Agrawal2019BenchmarkingOD} consisting of $133$ pairs with peptides' lengths ranging from 9 to 15 as the test set. We eliminate all complexes from \texttt{PPBench2024} whose PDB-ID are the same as those in \texttt{PPDBench} to avoid potential data leakage.
\subsection{Baseline Models}
Our method has been opened to the public in \url{https://github.com/Edapinenut/ppflow}. For comparison, we extend two models for the following tasks, including

(i) \textsc{DiffPP}, as a variant of \textsc{DiffAB}, which is a diffusion model for generating CDRs in antibodies targeted at antigens. \textsc{DiffPP} parametrizes the protein backbones in the same way as \textsc{AlphaFold2}, in which the positions of atoms in a residue are determined by C$\alpha$'s translation vector $\bm{\mathrm{x}}_{i,2}$ and the rotation matrix $O_i \in \mathrm{SO}(3)$ of the frame constructed by positions of $\{$N, C$\alpha$, C$\}$. The diffusion and reverse process on translation and orientation variables are modeled by \textsc{DDPM} \cite{ho2020denoising} and \textsc{SO(3)-DPM} \cite{leach2022denoising}.  For amino acid types, it uses a multinomial diffusion \cite{hoogeboom2021argmax}. 
 
(ii)  \textsc{DiffBP-PP}, as a variant of \textsc{DiffBP} or \textsc{TargetDiff}, as an atom-level diffusion model for generating molecules that bind to specific proteins.  It decomposes the residues into atoms, and uses DDPM to model the distribution of atoms' positions and D3PM \cite{austin2021structured} to predict amino acid types through mask language modeling techniques.\vspace{-0.3em}

These two models are the state-of-the-art models for target-specific protein and molecule generation, respectively. For other classical models, we describe them in different tasks.
\subsection{Peptide Generation}
\label{exp:gen}
\textbf{Metrics.} 
We choose 6 metrics for evaluating the quality of generated peptides. 
$\bm{\Delta G}$ is binding energies calculated by \textsc{ADCP} re-docking \cite{Zhang2019AutoDockCC}, since \textsc{ADCP} has shown the best performance in estimating the binding energies and poses of protein-peptide complexes \cite{Weng2020ComprehensiveEO}. {$\Delta G$} reflects the potential of a peptide being pharmaceutically active towards the targets, and \textbf{IMP\%-B} gives the percentages of the designed peptides with lower (better) $\Delta G$ than the reference peptides. The mean of reference $\Delta G$ is  $-427.72 \mathrm{kcal/mol}$. 
Stability evaluates binding scores of the original pose of the generated peptides, which directly reflects the quality of generated peptides without re-docking. The stability scores are calculated by \textsc{FoldX} \cite{Schymkowitz2005TheFW} since it performs fast and accurate stability calculation in protein binding tasks compared with other energy-based methods \cite{Luo2023RotamerDE}. We give \textbf{IMP\%-S} as the percentages of the designed ones with better stability than reference rather than average because some extremely unstable structures will make the comparison on average values meaningless. \textbf{Validity} is the ratio of the designed peptides that is chemically valid, through the criterion of whether the bonds of the atoms that should be bonded are broken. We follow our filtering process of datasets and define the bond is not broken if its length is within 0.5Å above and below the ideal value. \textbf{Novelty} is measured considering two aspects of structure and sequence: (i) the fraction of peptides with TM-score $< 0.5$ as used in \cite{lin2023generating} and (ii) the sequence overlap (SeqOL) less than 0.5. The given score is the mean ratio of novel peptides \emph{v.s.} references. \textbf{Diversity} is the  product of pairwise $(1 -$TM-score$)$ and $(1 - $SeqOL$)$ of the generated samples averaged across their target proteins. The higher the score, the more diverse peptides a model generates. Note that all the metrics except Validity are measured with chemically valid peptides. 

\textbf{Setup.}
For each model for comparison, we generate 20 peptides for each conditioned protein because the evaluation process requires excessive time. For example, $\Delta G$ obtained by \textsc{ADCP} requires more than 10 minutes for one pair of protein and peptides on a server with 128 CPU threads. Besides, since the greater the number of atoms, the more likely it is that binding interactions will occur, we eliminate the effects of peptide sizes (lengths) on the binding affinity by setting the generated sequence lengths the same as references.
\begin{figure*}[t]
    \begin{minipage}{.61\textwidth}
\centering
\captionof{table}{Comparison for flexible peptide re-docking.} \label{tab:dockcomp}\vspace{-0.8em}
\resizebox{1.0\columnwidth}{!}{
\begin{tabular}{lrrrrrr}
\toprule
         \begin{tabular}{lrrrrrrrrrc}
       & \multicolumn{3}{c}{L-RMSD$(\downarrow)$} & \multicolumn{3}{c}{C-RMSD$(\downarrow)$} & \multicolumn{3}{c}{Success\%$(\uparrow)$} & \multirow{2}{*}{\begin{tabular}[c]{@{}c@{}}Time\\ (s)\end{tabular}}  \\
       \cmidrule(lr){2-4} \cmidrule(lr){5-7} \cmidrule(lr){8-10}
      Methods & \multicolumn{1}{c}{10\%} & \multicolumn{1}{c}{30\%} & \multicolumn{1}{c}{50\%} & 
      \multicolumn{1}{c}{10\%} & \multicolumn{1}{c}{30\%} & \multicolumn{1}{c}{50\%} & \multicolumn{1}{c}{2} & \multicolumn{1}{c}{4} & \multicolumn{1}{c}{8} &             \\
      \midrule
\textsc{DiffPP} & \textbf{8.01}    & \textbf{15.26}   & 28.47  & 6.35   & 12.68   & 30.25   & \textbf{9.24}    & \textbf{20.21}    & \textbf{34.45}    & 27   \\
\textsc{PPFlow} & 12.44   & 17.82   & 31.24  & \textbf{2.60}   & \textbf{3.28}    & \textbf{4.26}    & 6.13    & 11.24    & 22.47    & 46   \\
\midrule
\textsc{Hdock}  & 9.95    & 18.65   & \textbf{23.07}  & 7.29   & 10.65   & 18.84   & 6.40    & 10.72    & 25.50    & 186  \\
\textsc{Vina}   & 11.38   & 19.03   & 27.69  & 6.27   & 9.56    & 16.59   & 5.45    & 9.96    & 21.12    & 1325
\end{tabular}    \\
\bottomrule
\end{tabular}
} 
\end{minipage}\vspace{-0.5em}
    \begin{minipage}{.38\textwidth}\centering
    \centering
\captionof{table}{Comparison for side-chain packing.} \vspace{-0.8em}
    \label{tab:sidechain}\resizebox{1.0\columnwidth}{!}{
\begin{tabular}{llrrrr}
\toprule
 & Methods        & $\chi_1$  & $\chi_2$  & $\chi_3$  & $\chi_4$  \\
 \midrule
\multirow{3}{*}{\rotatebox{90}{MAE}}                     & \textsc{Rosetta}        & 38.58 & \textbf{39.41} & 68.75 & 64.54 \\
                     & \textsc{RDE-PP}(w/o pt) & 44.28 & 51.22 & 70.21 & 71.85 \\
                     & \textsc{RDE-PP}        & \textbf{37.24} & 47.67 & \textbf{66.88} & \textbf{62.86} \\
                     \midrule
\multirow{2}{*}{\rotatebox{90}{NLL}} & \textsc{RDE-PP}(w/o pt) & 0.75  & 1.26  & 3.37  & 4.10  \\
                     & \textsc{RDE-PP}         & \textbf{0.62}  & \textbf{0.78}  & \textbf{3.05}  & \textbf{3.85} \\
                     \bottomrule
\end{tabular}}          
    \end{minipage}\vspace{-0.5em}
\end{figure*}

\textbf{Results.} Table.~\ref{tab:gencomp} gives the comparison of peptide generation task with generated examples shown in Figure.~\ref{fig:genvis}. For \textsc{PPFlow}, we test two variants of it. \textsc{PPFlow-BB} (PPFlow-BackBone) uses the generated backbone structures with \textsc{Rosetta} side-chain packing \cite{Alford2017TheRA} before re-docking scoring and stability calculation, and side-chain atoms in \textsc{PPFlow-FA} (PPFlow-FullAtom) is predicted by \textsc{RDE}. Note that the backbone structures before side-chain packing are the same for the two variants, leading to metrics except IMP\%-S being of little difference. Besides, for the others, the side-chain atoms are constructed by \textsc{Rosetta}.     It can be concluded that (i) \textsc{PPFlow} has the greatest potential to generate peptide drugs with high binding affinity towards the target protein according to $\Delta G$ and IMP\%-B metrics. (ii)  The complexes of peptides of original poses generated by \textsc{PPFlow} and the target proteins are the most stable, with the highest IMP\%-S metrics. (iii) \textsc{PPFlow} and \textsc{DiffPP} generate a comparably high ratio of novel peptides, while peptides generated by \textsc{PPFlow} are more diverse. (iv) The rotamers estimated by \textsc{RDE-PP} are more stable in the binding site than \textsc{Rosetta} according to IMP\%-S metrics of \textsc{PPFlow-BB} and \textsc{PPFlow-FA}. (v) With the bond lengths and angles fixed, the structures generated by \textsc{PPFlow} are chemically valid, while \textsc{DiffBP-PP} shows the lowest Validity due to the highest degree of freedom it models, \emph{i.e.} $4\times 3N_\mathrm{pp}$, for atom-level backbone structure, which also empirically shows the infeasibility of SBDD methods to generalize to peptide drug design. Therefore, in the following parts, we will not compare \textsc{DiffBP-PP} as baselines.


\subsection{Peptide Optimization}

\textbf{Setup.} We apply our model to another common pharmaceutical application: optimization of the existing peptides. To optimize a peptide, we first choose a start time $\mathrm{start\_time}\in [0,1]$, and use the probability paths constructed in Sec.~\ref{sec:torusflow}, ~\ref{sec:se3flow} and ~\ref{sec:typeflow}, to sample a perturbed peptide $\mathcal{L}_{t_{\mathrm{s}}} = \{X_{t_{\mathrm{s}}}^{(i)*}, s^{(i)}_{t_{\mathrm{s}}}\}_{i=1}^{N_\mathrm{pp}}$. By using the ODE sampling methods in Sec.~\ref{sec:sampling}, the trained \textsc{PPFlow} iteratively updates the sequences and structures to recover the peptides as a set of optimized ones.  For \textsc{DiffPP}, we use the same setting as \cite{luo2022antigenspecific}. The `$\mathrm{num\_step}$' is the number of how many optimization steps used in the generative diffusion process, where the start step of noised peptides is $(\mathrm{total\_step} - \mathrm{time\_step})$. The conversion between them is $\mathrm{total\_step} - \mathrm{num\_step} = \mathrm{total\_step} \times \mathrm{start\_time} $.   We sample 20 new peptides for each protein receptor in \texttt{PPDBench} and evaluate them in the same metrics as discussed in Sec.~\ref{exp:gen}. We set $\mathrm{total\_step}=100$ in this part. 

\textbf{Results.} Figure.~\ref{fig:optcomp} gives the comparison on peptide optimization results.
It can be concluded that (i) The larger number of optimization steps contributes little to the improvement in the binding affinity, but \textsc{PPFlow} usually generates more peptides with higher binding affinities; (ii) Optimized peptides are usually similar to the original one in small optimization steps since the novelty are smaller, which is desired in many practical applications;
(iii) The diversity is more controllable in \textsc{PPFlow} by changing the number of optimization steps.
\subsection{Protein-Peptide Docking}
Besides the two tasks, in this part, we generalize our model to the peptide flexible re-docking task to figure out if the model can learn the binding poses of the ligands. The task can be regarded as establishing a probabilistic model of $p(\{X^{(i)}\}_{i=0}^{N_{\mathrm{pp}}}| \{s^{(i)}\}_{i=0}^{N_{\mathrm{pp}}}, \mathcal{R}\})$. For \textsc{DiffPP} and \textsc{PPFlow}, $\{X_0^{(i)}\}$, as the initialized peptide structures, are obtained by first moving them to the pocket center, and then adding Gaussian noise to each atom. We retrain \textsc{DiffPP} and \textsc{PPFlow} with the sequences given as additional conditions. For classical docking methods, we choose \textsc{HDock} \cite{Yan2020Hdock} which is a re-docking method for protein-protein interactions, and \textsc{VinaDock} \cite{Eberhardt2021AutoDockV1} proposed for molecule docking. We choose the best 10\%/30\%/50\% poses for comparison on ligand-RMSD (\textbf{L-RMSD}) between C$\alpha$ atoms and centroid-RMSD (\textbf{C-RMSD}), and the success rate (\textbf{Success\%}) of the docked ligands' RMSD smaller than 2/4/8Å. It can be concluded in Table.~\ref{tab:dockcomp} that (i) \textsc{DiffPP} shows the best docking performance on L-RMSD and Success\% because it directly optimized the RMSD between C$\alpha$ atoms, and \textsc{PPFlow} models the ligands' centroid most accurately.  (ii) The deep-learning-based generative models can achieve competitive performance with the classical energy-based ones, while the latter requires enormous rounds of iteration, costing dozens of times longer for computation.
\vspace{-0.3em}
\subsection{Side-Chain Packing}\label{sec:exp:sidechain}
\vspace{-0.2em}
To figure out the effectiveness of our fine-tuned model of \textsc{RDE-PP}, We compare it with two baseline methods Rosetta(fixbb) \cite{KoehlerLeman2019MacromolecularMA} and the unpretrained version \textsc{RDE-PP}(w/o pt). For each reference peptide, we initialize the rotamers randomly with uniform angle distribution, and sample 20 conformations on side chains. We evaluate the mean absolute error (\textbf{MAE}) of the predicted sidechain torsional angles as $\mathrm{MAE} = |(\hat\chi - \chi)\bmod{(2\pi)}|$ where $\hat\chi$ is the estimated ones and $\chi$ is the ground truth, and negative likelihood (\textbf{NLL}). In Table.~\ref{tab:sidechain}, the results demonstrate that the RDE-PP with pretraining outperforms the baselines on three of the four torsional angles in terms of the evaluated metrics.
\vspace{-0.2em}
\section{Conclusion}
\vspace{-0.3em}
In this paper, we focus on the target-specific peptide design tasks. To fulfill it, we first establish a dataset called \texttt{PPBench2024}, and then propose a flow-matching generative model on torus manifolds called \textsc{PPFlow}, which attempts to learn the distribution of torsion angles of peptide backbones directly.
Experiments are conducted to evaluate its performance on several tasks, indicating its superior performance on target-aware peptide design and optimization.

However, since the classical docking methods for binding pose calculation are extremely slow,  we extend our model to docking tasks. Still, it is not as competitive as the baseline models, which will be our future focus. Besides, the performance gaps between the extended side-chain packing model and the classical ones are still small, urging us to develop a side-chain packing model with high prediction accuracy. 

\section*{Acknowledgements}
This work was supported by the Science \& Technology Innovation 2030 Major Program Project No. 2021ZD0150100, National Natural Science Foundation of China Project No. U21A20427, Project No. WU2022A009 from the Center of Synthetic Biology and Integrated Bioengineering of Westlake University, and Project No. WU2023C019 from the Westlake University Industries of the Future Research. Finally, we thank the Westlake University HPC Center for providing computational resources. Besides, we thank the help of Dr.~Tailin Wu, and his great efforts in his deep insights into cutting-edge issues, the guidance provided in the rebuttal, and the funding that supplied us with the necessary equipment for our research.

\section*{Impact Statement}
With the long-term continuation of COVID-19, more and more AI scientists are beginning to turn their research interests to drug design. The focus of this paper is on one of them -- peptide drugs. There is very limited work aimed at using AI algorithms to design and optimize peptide drugs. However, peptide drugs are now rising as effective therapeutics, with hundreds of them proving to be successful, such as semaglutide for obesity and diabetes. Here, this work presents a considerable contribution to establishing a large dataset and deep-learning model for protein-specific peptide design, in which the effectiveness is validated empirically. As far as we know, it is the first AI-assisted peptide drug design solution, so we here emphasize the social significance of the work and hope to get the attention of the reviewers and the review committee.

\bibliography{example_paper}
\bibliographystyle{icml2023}

\newpage
\appendix
\onecolumn
\section{Method}
\subsection{Proof of Proposition.~\ref{prop1}}
\label{app:proof1}
By Equation.~\ref{eq:wnpath}, we can obtain  $\bm{\epsilon} = \frac{ \bm{\tau}_t - \bm{\mu}_t}{\sigma_t}$. Besides, from the definition of $u_t(\bm{\tau}|\bm{\tau}_1) = d\bm{\dot \tau}/ dt$, it can be written as 
\begin{equation}
\begin{aligned}
u_t(\bm{\tau}|\bm{\tau}_1) =& \dot \sigma_t\bm{\epsilon} + \bm{\dot\mu}_t \\
  =&  \dot \sigma_t(\frac{ \bm{\tau}_t - \bm{\mu}_t}{\sigma_t}) + \bm{\dot\mu}_t \\
  =&   \frac{\dot\sigma_t(\bm{\tau}_1)}{\sigma_t(\bm{\tau}_1)} \left(\bm{\tau} - \bm{\mu}_t(\bm{\tau}_1)\right) +  \bm{\dot\mu}_t(\bm{\tau}_1).
\end{aligned}
\end{equation}
\subsection{Proof of Proposition.~\ref{prop2}}
\label{app:proof2}
Firstly, we claim the disintegration of measures of $p_t(\bm{\tau})$, as $p_t(\bm{\tau}) = \prod_i p(\tau^{(i)})$.
For $p_0(\bm{\tau})$, the disintegration satisfies. For $p_1(\bm{\tau})$, our parametrization assumes the torsion angles are orthogonal and the distribution is independent, so it also satisfies disintegration.
In this way, it is easy to obtain an intermediate probability $p_t$ satisfies disintegration, and similar for the conditional probability $p_t(\bm{\tau}|\bm{\tau}_1)$.

Then we can factorizes the metric on $\mathbb{T}^N$ into $\mathbb{S} \times \cdots \times \mathbb{S}$, and $p(\tau) \in \mathcal{P}(\mathbb{S})$.

Let $u_t = \mathbb{E}_{\tau_0\sim p_0, \tau_1 \sim p_1}\left[\frac{p_t(\tau|\tau_0,\tau_1)}{p_t(\tau)}u_t(\tau|\tau_0,\tau_1)\right]$, and
\begin{equation}
\begin{aligned}
    &\nabla_\theta (\mathbb{E}_{\tau_0\sim p_0, \tau_1 \sim p_1, \tau\sim p_t(\tau| \tau_0,\tau_1)}[\|v_t(\tau) - u_t(\tau|\tau_0,\tau_1)\|^2] - \mathbb{E}_{\tau\sim p_t}[\|v_t(\tau) - u_t(\tau)\|^2])\\
  =&-2\nabla_\theta (\mathbb{E}_{\tau_0\sim p_0, \tau_1 \sim p_1, \tau\sim p_t(\tau| \tau_0,\tau_1)}\langle  v_t(\tau), u_t(\tau|\tau_0,\tau_1)\rangle  - \mathbb{E}_{\tau\sim p_t}\langle v_t(\tau) , u_t(\tau)\rangle ).
  \end{aligned}\label{eq:gradient}
\end{equation}
Then,
\begin{equation}
    \begin{aligned}
        &\mathbb{E}_{\tau\sim p_t}\langle  v_t(\tau), u_t(\tau)\rangle \\
        =&\int \langle v_t(\tau), u_t(\tau)\rangle  p_t(\tau) d\tau \\
        =& \int \langle v_t(\tau),\mathbb{E}_{p_0,p_1}\left[\frac{p_t(\tau|\tau_0,\tau_1)}{p_t(\tau)}u_t(\tau|\tau_0,\tau_1)\right]\rangle  p_t(\tau) d\tau \\
        =& \int \langle v_t(\tau),\int \frac{p_t(\tau|\tau_0,\tau_1)}{p_t(\tau)}u_t(\tau|\tau_0,\tau_1) p_0(\tau_0)p_1(\tau_1)d\tau_0 \tau_1 \rangle  p_t(\tau) d\tau\\
        =& \int \int \langle v_t(\tau),  u_t(\tau|\tau_0, \tau_1)  \rangle p_t(\tau|\tau_0,\tau_1) p_0(\tau_0)p_1(\tau_1) d\tau d\tau_0 d\tau_1\\
        =&\mathbb{E}_{\tau_0\sim p_0, \tau_1 \sim p_1, \tau\sim p_t(\tau| \tau_0,\tau_1)} \langle v_t(\tau), u_t(\tau)\rangle,
    \end{aligned}
\end{equation}
where in the last equality we change the order of integration. It proves Equation.~\ref{eq:gradient} equals 0.

\subsection{Proof of Proposition.~\ref{prop3}}
\label{app:proof3}
The proof is inspired by \cite{kohler20a, lin2023d3fg}, as follows:

\textbf{Lemma A1.} Let $\mathrm{T}_g(\cdot)$ be the operation in $\mathrm{SE}(3)$, if the following update function in the ODE sampler (Sec.~\ref{sec:sampling}) for the atom level's positions are defined as 
\begin{equation}
\begin{aligned}
        v_t(\{X^{(i)*}\}| \mathcal{C}_t^*) &= (\{X_{t+\Delta t}^{(i)*}\} - \{X_{t}^{(i)*}\}) / \Delta t,
\end{aligned}
\end{equation}
in which $\{X_{t}^{(i)*}\} = O_t \mathrm{nerf}(\bm{\tau}_t)+ \bm{\mathrm{x}}^{\mathrm{(C)}}$. 
The invariance and equivariance of the following functions in the updating process  are listed as 
\begin{equation}
\begin{aligned}
    v\left(\{X^{(i)}_t\}|\mathrm{T}_g(\mathcal{C}^{*})\right)
   &= \mathrm{T}_g\left( v(\{X^{(i)}_t\}|\mathcal{C}^{*})\right); \\
   v\left((\{s^{(i)}_t\})|\mathrm{T}_g(\mathcal{C}^{*})\right)
   &= v\left((\{s^{(i)}_t\})|\mathcal{C}^{*})\right) ,
\end{aligned}
\end{equation}
and the prior distribution as 
\begin{align} 
    p_0(\{X^{(i)*}_0\} | \mathrm{T}_g(\mathcal{R})) &=  p_0(\{X^{(i)*}_0\}|\mathcal{R})\\
    p_0(\{s^{(i)*}_0\} | \mathrm{T}_g(\mathcal{R})) &=  p_0(\{s^{(i)*}_0\}|\mathcal{R}).
\end{align}
Then the distribution which the final structures and sequences are sampled from as $p\left(\mathcal{L}|\mathcal{R}\right)$ is SE(3)-equivariant.

\textbf{Proof:}
In the following, we write $\bm{X} = \{X^{(i)*}\}_{i=1}^{N_{\mathrm{pp}}}$, and $\bm{s} = \{s^{(i)}\}_{i=1}^{N_{\mathrm{pp}}}$ for notation simplicity. For the defined $v_t$, we first obtain that the update process is equivariant, as 
\begin{equation}
\begin{aligned}
    &v_t(\bm{X}| \mathrm{T}_g(\mathcal{C}_t^*)) \Delta t + \mathrm{T}_g(\bm{X}_t) \\
   =&\mathrm{T}_g\left( v_t(\bm{X}|\mathcal{C}_t^{*})\right)\Delta t   + \mathrm{T}_g(\bm{X}_t)\\
   =& \mathrm{T}_g\left( v_t(\bm{X}|\mathcal{C}_t^{*}) \Delta t  +\bm{X}_t \right)\\
   =& \mathrm{T}_g\left(\bm{X}_{t+ \Delta t} \right)
\end{aligned}
\end{equation}
and for each transition kernel $p(\bm{X}_{t+\Delta t} | \mathcal{C}_t^*)$, it is SE(3)-equivariant, since
\begin{align}
    &p(\mathrm{T}_g(\bm{X}_{t+ \Delta t}) | \mathrm{T}_g({\mathcal{C}_t^*})) \\
   =&p(\mathrm{T}_g(\bm{X}_t + v_t(\bm{X}| \mathrm{T}_g({\mathcal{C}_t^*})) \Delta t   )| \mathrm{T}_g(\mathcal{C}_t^*))\\
   =&\int p(\mathrm{T}_g(\bm{X}_{t+ \Delta t}) - \mathrm{T}_g(\bm{X}_t),  \mathrm{T}_g(\bm{X}_t))| \mathrm{T}_g({\mathcal{C}_t^*})) d\mathrm{T}_g(\bm{X}_{t}) \\
   =&\int p(\mathrm{T}_g(\bm{X}_{t+ \Delta t}) - \mathrm{T}_g(\bm{X}_t)|\mathrm{T}_g({\mathcal{C}_t^*})) p(\mathrm{T}_g(\bm{X}_t) |\mathrm{T}_g({\mathcal{C}_t^*}) )
   d\bm{X}_{t}\\
   =& \int p(v_t(\bm{X}| \mathrm{T}_g(\mathcal{C}_t^*)) \Delta t|\mathrm{T}_g({\mathcal{C}_t^*})) p(\bm{X}_t |{\mathcal{C}_t^*})  d\bm{X}_{t}\\
   =& \int p(v_t(\bm{X}|\mathcal{C}_t^*)\Delta t|{\mathcal{C}_t^*}) p(\bm{X}_t |{\mathcal{C}_t^*})  d\bm{X}_{t}\\
   =& \int p(\bm{X}_{t+\Delta t} -\bm{X}_{t}\} |{\mathcal{C}_t^*}) p(\bm{X}_{t} |{\mathcal{C}_t^*})  d\bm{X}_{t}\\
   =& p(\bm{X}_{t+\Delta t} | {\mathcal{C}_t^*})
\end{align}
And it is easy to obtained that the transition kernel $p(\bm{s}_{t+\Delta t}|\mathcal{C}^*_t)$ is SE(3)-invariant, since $v_t(\bm{s}|{\mathcal{C}_t^*})$ is invariant.
Besides, for $p_0(\{X_0^{(i)*}\})$ and $p_0(\{s_0^{(i)*}\})$, it is a SE(3)-invariant distribution.
Therefore,
\begin{align}
    &p\left(\mathrm{T}_g(\bm{X}_1),\bm{s}_{1}|\mathrm{T}_g(\mathcal{R})\right)\\
    =& \int p\left(\mathrm{T}_g(\bm{X}_{0}),\bm{s}_{0}|\mathrm{T}_g(\mathcal{R})\right)
    \prod_{k=1}^{K} p\left(\mathrm{T}_g(\bm{X}_{k\Delta t}),\bm{s}_{k\Delta t}|\mathrm{T}_g(\mathcal{C}_{(k-1)\Delta t}^*)\right) d\mathcal{L}_{0:(K-1)\Delta t}\\
    =& \int p\left(\bm{X}_{0},\bm{s}_{0}|\mathcal{R}\right)
    \prod_{k=1}^{K} p\left(\mathrm{T}_g(\bm{X}_{k\Delta t})|\mathrm{T}_g(\mathcal{C}_{(k-1)\Delta t}^*)\right)p\left(\bm{s}_{k\Delta t}|\mathrm{T}_g(\mathcal{C}_{(k-1)\Delta t}^*)\right) d\mathcal{L}_{0:(K-1)\Delta t}\\
    =& \int p\left(\bm{X}_{0},\bm{s}_{0}|\mathcal{R}\right)
    \prod_{k=1}^{K} p\left(\bm{X}_{k\Delta t}|\mathcal{C}_{(k-1)\Delta t}^*\right)p\left(\bm{s}_{k\Delta t}|\mathcal{C}_{(k-1)\Delta t}^*\right) d\mathcal{L}_{0:(K-1)\Delta t}\\
    =& \int p\left(\bm{X}_{0},\bm{s}_{0}|\mathcal{R}\right)
    \prod_{k=1}^{K} p\left(\bm{X}_{k\Delta t},\bm{s}_{k\Delta t}|\mathcal{C}_{(k-1)\Delta t}^*\right) d\mathcal{L}_{0:(K-1)\Delta t}\\
    =&p\left(\bm{X}_1,\bm{s}_{1}|\mathcal{R}\right),
\end{align}
where $K$ is the step number of the ODE sampler, and $\Delta t = 1/K$. Because $p\left(\mathrm{T}_g(\bm{X}_1),\bm{s}_{1}|\mathrm{T}_g(\mathcal{R})\right) = p\left(\bm{X}_1,\bm{s}_{1}|\mathcal{R}\right)$, which is equivalent to $p\left(\mathrm{T}_g(\mathcal{L}_1^*)|\mathrm{T}_g(\mathcal{R})\right) = p\left(\mathcal{L}_1^*|\mathcal{R}\right)$, the Lemma is proved.

Then we write $\{X_{t}^{(i)*}\} = O_t \mathrm{nerf}(\bm{\tau}_t)+ \bm{\mathrm{x}}_t^{\mathrm{(C)}}$, and the following proposition can give the conditions that the transition kernels of $p(O_{t+\Delta t}|\mathcal{C}^*_t)$, $p(\bm{\mathrm{x}}^{\mathrm{C}}_{t+\Delta t}|\mathcal{C}_t^*)$ and $p(\bm{\tau}_{t+\Delta t}|\mathcal{C}_t^*)$ and $p(\bm{s}_{t+\Delta t}|\mathcal{C}_t^*)$ should satisfiy.

\textbf{Lemma A2.} If $p(O_{0}^{\mathrm{(C)}})$, $p(\bm{\mathrm{x}}^{\mathrm{(C)}}_{0})$, $p(\bm{\tau}_{0})$ and $p(\bm{s}_0)$ are SE(3)-invariant, and $p(O^{\mathrm{(C)}}_{t+\Delta t}|\mathcal{C}_t)$ is SO(3)-equivariant and T(3)-invariant, $p(\bm{\mathrm{x}}^{\mathrm{C}}_{t+\Delta t}|\mathcal{C}_t^*)$ is SE(3)-equivariant, $p(\bm{\tau}_{t+\Delta t}|\mathcal{C}_t^*)$ and $p(\bm{s}_{t+\Delta t}|\mathcal{C}_t^*)$ is SE(3)-invariant, then $p\left(\mathrm{T}_g(\mathcal{L}_1^*)|\mathrm{T}_g(\mathcal{R})\right) = p\left(\mathcal{L}_1^*|\mathcal{R}\right)$, where $\mathcal{L}_1^* = (O_1^{\mathrm{(C)}} \mathrm{nerf}(\bm{\tau}_1)+ \bm{\mathrm{x}}_1^{\mathrm{(C)}}, \bm{s}_1) $.

\textbf{Proof:}
Here for $\mathrm{T}_g$, we can decompose it as $\mathrm{T}_g = \mathrm{T}_r \circ \mathrm{T}_t$, meaning the rotation and translation opeartions as $\mathrm{SE}(3) \cong \mathrm{SO}(3) + \mathrm{T}(3)$. By this mean,
\begin{align}
     &p\left(\mathrm{T}_g(\mathcal{L}_1^*)|\mathrm{T}_g(\mathcal{R})\right) \\
    =&p\left(\mathrm{T}_g(O_1^{\mathrm{(C)}} \mathrm{nerf}(\bm{\tau}_1)+ \bm{\mathrm{x}}_1^{\mathrm{(C)}}), \bm{s}_1)|\mathrm{T}_g(\mathcal{R})\right)\\
    =&p\left(\mathrm{T}_g(O_1^{\mathrm{(C)}} \mathrm{nerf}(\bm{\tau}_1)) + \mathrm{T}_g(\bm{\mathrm{x}}_1^{\mathrm{(C)}})), \bm{s}_1|\mathrm{T}_g(\mathcal{R})\right)\\
    =&p\left(\mathrm{T}_g(O_1^{\mathrm{(C)}} \mathrm{nerf}(\bm{\tau}_1)) |\mathrm{T}_g(\mathcal{R})\right) p\left( \mathrm{T}_g(\bm{\mathrm{x}}_1^{\mathrm{(C)}}))|\mathrm{T}_g(\mathcal{R})\right)p\left(\bm{s}_1|\mathrm{T}_g(\mathcal{R})\right)\\
    =&p\left(\mathrm{T}_g(O_1^{\mathrm{(C)}} \mathrm{nerf}(\bm{\tau}_1)) |\mathrm{T}_g(\mathcal{R})\right) p\left( \mathrm{T}_g(\bm{\mathrm{x}}_1^{\mathrm{(C)}}))|\mathrm{T}_g(\mathcal{R})\right)p\left(\bm{s}_1|\mathrm{T}_g(\mathcal{R})\right).
\end{align}
Because $\mathrm{nerf}(\bm{\tau}_1)$ is always reconstructed with unit rotation $\mathrm{diag}(1,1,1)$ and zero-mass-centered, therefore, 
\begin{align}
     &p\left(\mathrm{T}_g(O_1^{\mathrm{(C)}} \mathrm{nerf}(\bm{\tau}_1)) |\mathrm{T}_g(\mathcal{R})\right)\\
    =&p\left(\mathrm{T}_g(O_1^{\mathrm{(C)}})  \mathrm{nerf}(\bm{\tau}_1) |\mathrm{T}_g(\mathcal{R})\right)\\
    =&p\left(\mathrm{T}_r \circ \mathrm{T}_t(O_1^{\mathrm{(C)}})  \mathrm{nerf}(\bm{\tau}_1) |\mathrm{T}_g(\mathcal{R})\right) \\
    =&p\left(\mathrm{T}_r (O_1^{\mathrm{(C)}})  \mathrm{nerf}(\bm{\tau}_1) |\mathrm{T}_g(\mathcal{R})\right) \\
    =&\int p\left(\mathrm{T}_r (O_0^{\mathrm{(C)}})  \mathrm{nerf}(\bm{\tau}_0) |\mathrm{T}_g(\mathcal{R})\right) \prod_{k=1}^{K} p\left(\mathrm{T}_r({O}^{\mathrm{(C)}}_{k\Delta t}) \mathrm{nerf}(\bm{\tau}_{k\Delta t})|\mathrm{T}_g(\mathcal{C}_{(k-1)\Delta t}^*)\right) dM_{0:(K-1)\Delta t}\\
    =& \int p\left(O_0^{\mathrm{(C)}}  \mathrm{nerf}(\bm{\tau}_0) |\mathrm{T}_g(\mathcal{R})\right) \prod_{k=1}^{K} p\left({O}^{\mathrm{(C)}}_{k\Delta t} \mathrm{nerf}(\bm{\tau}_{k\Delta t})|\mathcal{C}_{(k-1)\Delta t}^*\right) dM_{0:(K-1)\Delta t}\\
    =&p\left(O^{\mathrm{(C)}}_1 \mathrm{nerf}(\bm{\tau}_1) |\mathcal{R}\right)
\end{align}
where $dM_{0:(K-1)\Delta t} = dO^{\mathrm{(C)}}_{0:(K-1)\Delta t} d\bm{\tau}_{0:(K-1)\Delta t}$. For $p\left( \mathrm{T}_g(\bm{\mathrm{x}}_1^{\mathrm{(C)}}))\right)$ and $p\left(\bm{s}_1|\mathrm{T}_g(\mathcal{R})\right)$, the equivariance and invariance are also easy to obtain. Therefore, the Lemma A2. is proved.

Finally, \textbf{PROOF OF PROPOSITION~\ref{prop3}}:

 For the transition kernels which are generated by the following equations as 
\begin{equation}
    \begin{aligned}
        p(\bm{\tau}_{t+\Delta t}|\mathcal{C}_t^*) &= p(\bm{\tau}_{t} + v_t(\bm{\tau}|\mathcal{C}_t^*)\Delta t|\mathcal{C}_t^*);\\
        p(\bm{\mathrm{x}}_{t+\Delta t}^{\mathrm{(C)}}|\mathcal{C}_t^*) &= p(\bm{\mathrm{x}}_{t}^{\mathrm{(C)}}+ v_t(\bm{\mathrm{x}}^{\mathrm{(C)}}|\mathcal{C}_t^*)\Delta t|\mathcal{C}_t^*);\\
        p(O^{\mathrm{(C)}}_{t+\Delta t}|\mathcal{C}_t^*) &= p(O^{\mathrm{(C)}}_{t} \exp_{O^{\mathrm{(C)}}_t}(v_t(\bm{\tau}|\mathcal{C}_t^*)\Delta t)|\mathcal{C}_t^*);\\
        p(\bm{s}_{t+\Delta t}|\mathcal{C}_t^*) &= p(\bm{s}_{t} + v_t(\bm{s}|\mathcal{C}_t^*)\Delta t|\mathcal{C}_t^*),\\
    \end{aligned}
\end{equation}
if the conditions of flow-matching vector fields in \textbf{Propostion~\ref{prop3}} are all satisfied, then $p(\bm{\tau}_{t+\Delta t}|\mathcal{C}_t^*)$,
$p(\bm{\mathrm{x}}_{t+\Delta t}^{\mathrm{(C)}}|\mathcal{C}_t^*)$, $p(O_{t+\Delta t}|\mathcal{C}_t^*)$ and $p(\bm{s}_{t+\Delta t}|\mathcal{C}_t^*)$ satisfies the conditions in \textbf{Lemma A.2} resepectively, and $v(\bm{x}|\mathcal{C}_t^{*})$ satisfies \textbf{Lemma A.1}. From both perspectives, the roto-translational equivariance can be proved.

We demonstarte the first proof path, to shown that the conditions of \textbf{Lemma A.2} holds.

For $\bm{\tau}$:
\begin{equation}
\begin{aligned}
    p(\mathrm{T}_g(\bm{\tau}_{t+\Delta t})|\mathrm{T}_g(\mathcal{C}_t^*)) &= p(\mathrm{T}_g(\bm{\tau}_{t} + v_t(\bm{\tau}|\mathcal{C}_t^*)\Delta t)|\mathrm{T}_g(\mathcal{C}_t^*))\\
    &=p(\mathrm{T}_g(\bm{\tau}_{t}) + \mathrm{T}_g(v_t(\bm{\tau}|\mathcal{C}_t^*)\Delta t)|\mathrm{T}_g(\mathcal{C}_t^*))\\
    &=\int p(\mathrm{T}_g(\bm{\tau}_{t+\Delta t}) -\mathrm{T}_g(\bm{\tau}_{t}) |\mathrm{T}_g(\mathcal{C}_t^*))   p(\mathrm{T}_g(\bm{\tau}_{t})|\mathrm{T}_g(\mathcal{C}_t^*)) d\mathrm{T}_g(\bm{\tau}_{t})\\
    &=\int p(\mathrm{T}_g(v_t(\bm{\tau}|\mathcal{C}_t^*)\Delta t)|\mathrm{T}_g(\mathcal{C}_t^*)) p(\mathrm{T}_g(\bm{\tau}_{t})|\mathrm{T}_g(\mathcal{C}_t^*)) d\mathrm{T}_g(\bm{\tau}_{t}) \\
    &=\int p(v_t(\bm{\tau}|\mathcal{C}_{t}^*)\Delta t|\mathcal{C}_t^*) p(\bm{\tau}_{t}|\mathcal{C}_t^*) d\bm{\tau}_{t} \\
    &=\int p(\bm{\tau}_{t+\Delta t} - \bm{\tau}_{t}|\mathcal{C}_t^*) p(\bm{\tau}_{t}|\mathcal{C}_t^*) d\bm{\tau}_{t} \\
    &= p(\bm{\tau}_{t+\Delta t}|\mathcal{C}_t^*)
\end{aligned}
\end{equation}
Beside, for the other three variables, the equations can be deduced similarly.

\subsection{\textsc{LoCS} Updating}
\label{app:locs}
Because the outputs should include the gradient vectors of rotation angle, global translational vector, rotation matrix, and type probability, we employ the \textsc{LoCS} to output the vector fields by 
\begin{equation}
    \begin{aligned}
        v_t(\bm{c}^{(i)}|\mathcal{C}^*) &= \mathrm{MLP}_{s}(\bm{h}_i);\\
        v_t(\{\phi^{(i)},\psi^{(i)},\omega^{(i)}\}|\mathcal{C}^*) &= \mathrm{MLP}_{\tau}(\bm{h}_i);\\
        v_t(\bm{\mathrm{x}}^{(\mathrm{C})}|\mathcal{C}^*) &= \mathrm{MLP}_{\rm{x}}(\bm{h}_\mathrm{G})O_t ;\\
        v_t(O^{\mathrm{(C)}}|\mathcal{C}^*) &= \mathrm{tran}_{O_t}(\mathrm{MLP}_{O}(\bm{h}_\mathrm{G})), \label{eq:nnout}
    \end{aligned}
\end{equation}
in which $\bm{h}_\mathrm{G} = \sum_i\bm{h}_i$ is the global representation obtained by summation, $\mathrm{MLP}_{s}:\mathbb{R}^{D} \rightarrow \mathbb{R}^{20}$, $\mathrm{MLP}_{\tau}:\mathbb{R}^{D} \rightarrow \mathbb{T}^{3}$, $\mathrm{MLP}_{\rm{x}}: \mathbb{R}^{D} \rightarrow \mathbb{R}^3$ and $\mathrm{MLP}_{\rm{x}}: \mathbb{R}^{D} \rightarrow \mathbb{R}^3$. $\mathrm{MLP}_{O}$ predicts a vector in Lie group $\mathfrak{so}(3)$, and translate it to $O_t$'s tangent space by $\mathrm{tran}_{O_t}(\cdot)$. The output vector fields satisfy the equivariance and invariance conditions in Proposition.~\ref{prop3}.

\section{Dataset Statistics}

\subsection{Data preprocess}
\label{app:dataprepro}
\begin{figure*}\centering
    \includegraphics[width=0.7\linewidth]{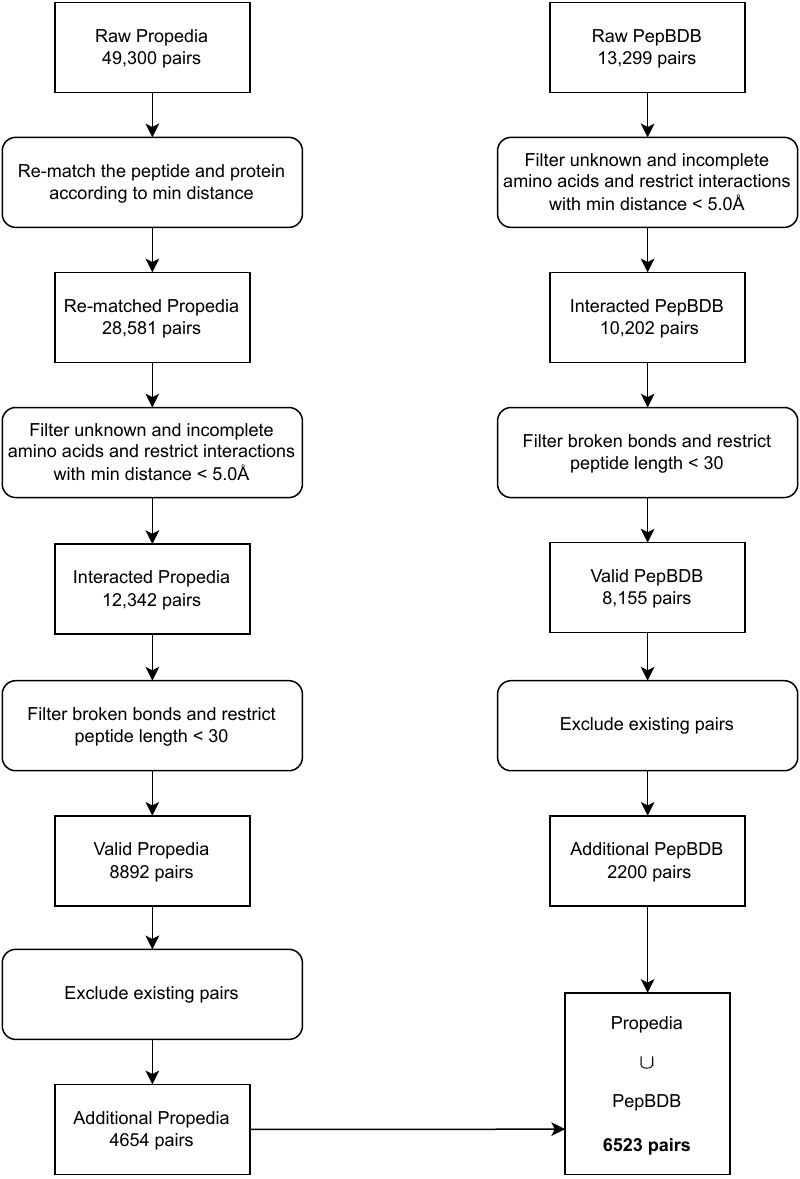}
        \caption{The process of screening the two datasets to expand the raw \texttt{PPBench}.}\label{fig:screen}
\end{figure*}

The construction of raw \texttt{PPBench2024} is given in the Sec.~\ref{sec:data}. For the additional complexes from \texttt{PropediaV2.3} and \texttt{PepBDB}, we give the detailed screening process in the Figure.~\ref{fig:screen}. Because in PropediaV2.3, one peptide can be paired with several chains in a protein complex, another re-matching step should be conducted first.

\begin{figure}
    \centering
    \subfigure[Histogram of peptide lengths]{\label{fig:pplength}\includegraphics[width=0.4\columnwidth]{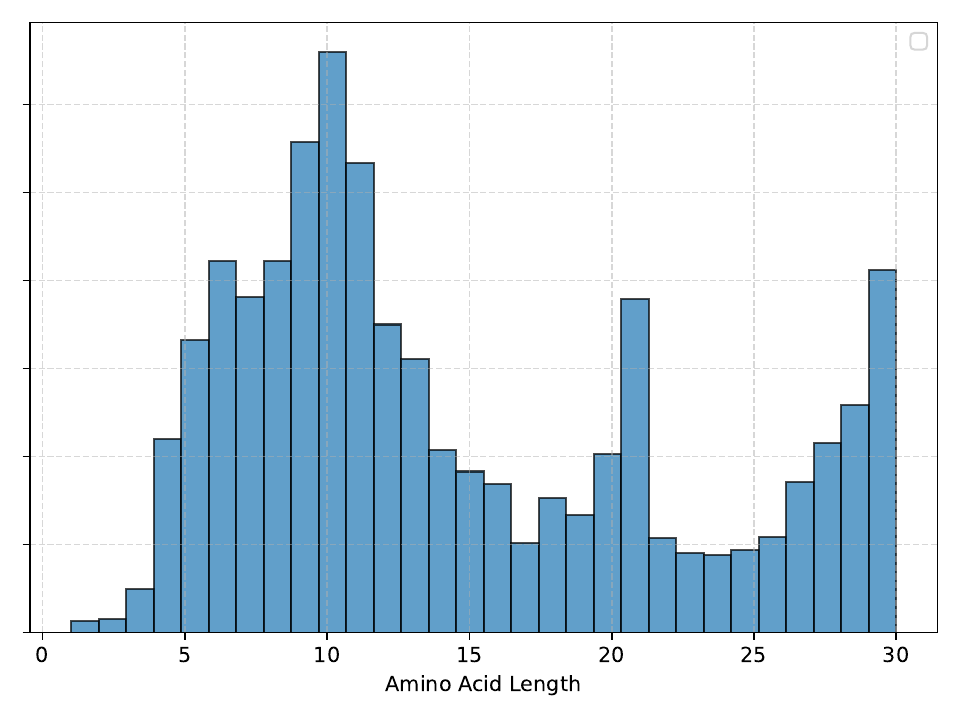}}
    \subfigure[Empirical and estimated bond length distribution.]{ \label{fig:emplendist}
    \includegraphics[width=1\linewidth, trim = 0 00 0 00,clip]{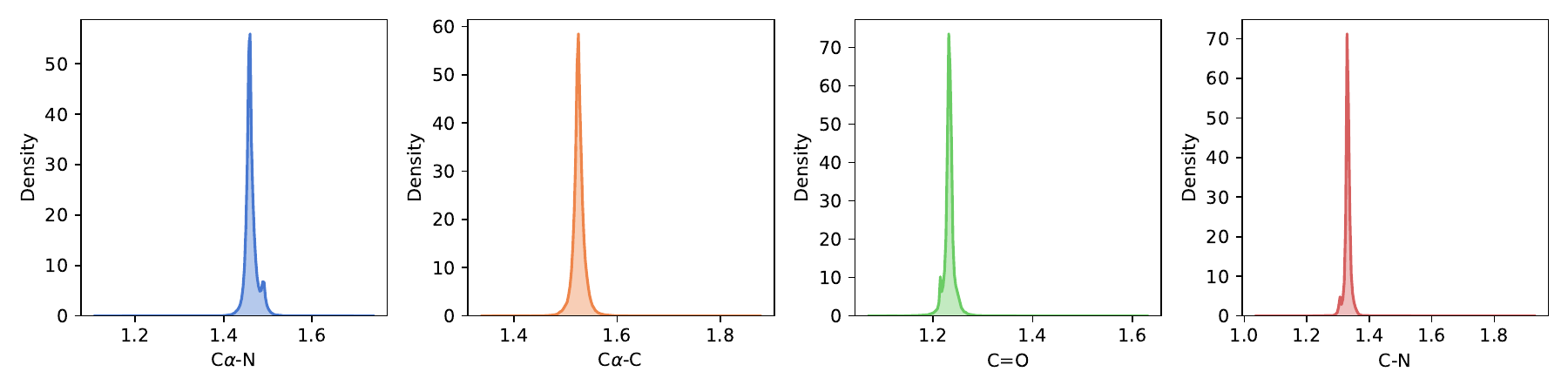}
    }\vspace{-0.5em}
    \subfigure[Empirical and estimated bond angle distribution.]{ \label{fig:empangdist} 
    \includegraphics[width=1\linewidth, trim = 0 100 0 100,clip]{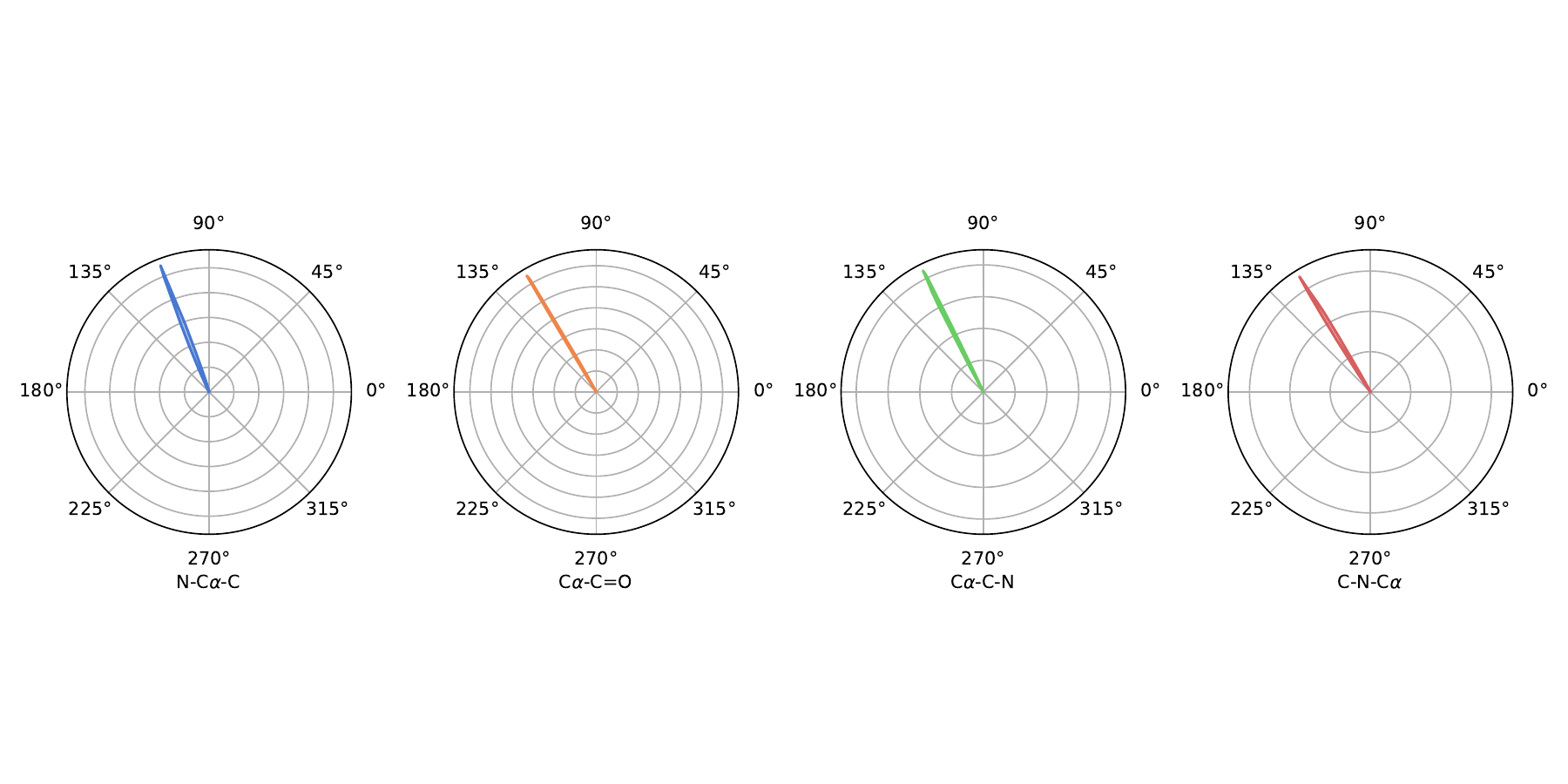}
    }\vspace{-0.5em}
    \subfigure[Empirical and estimated torsion angle distribution.]{ \label{fig:emptordist} 
    \includegraphics[width=1\linewidth, trim = 0 100 0 100,clip]{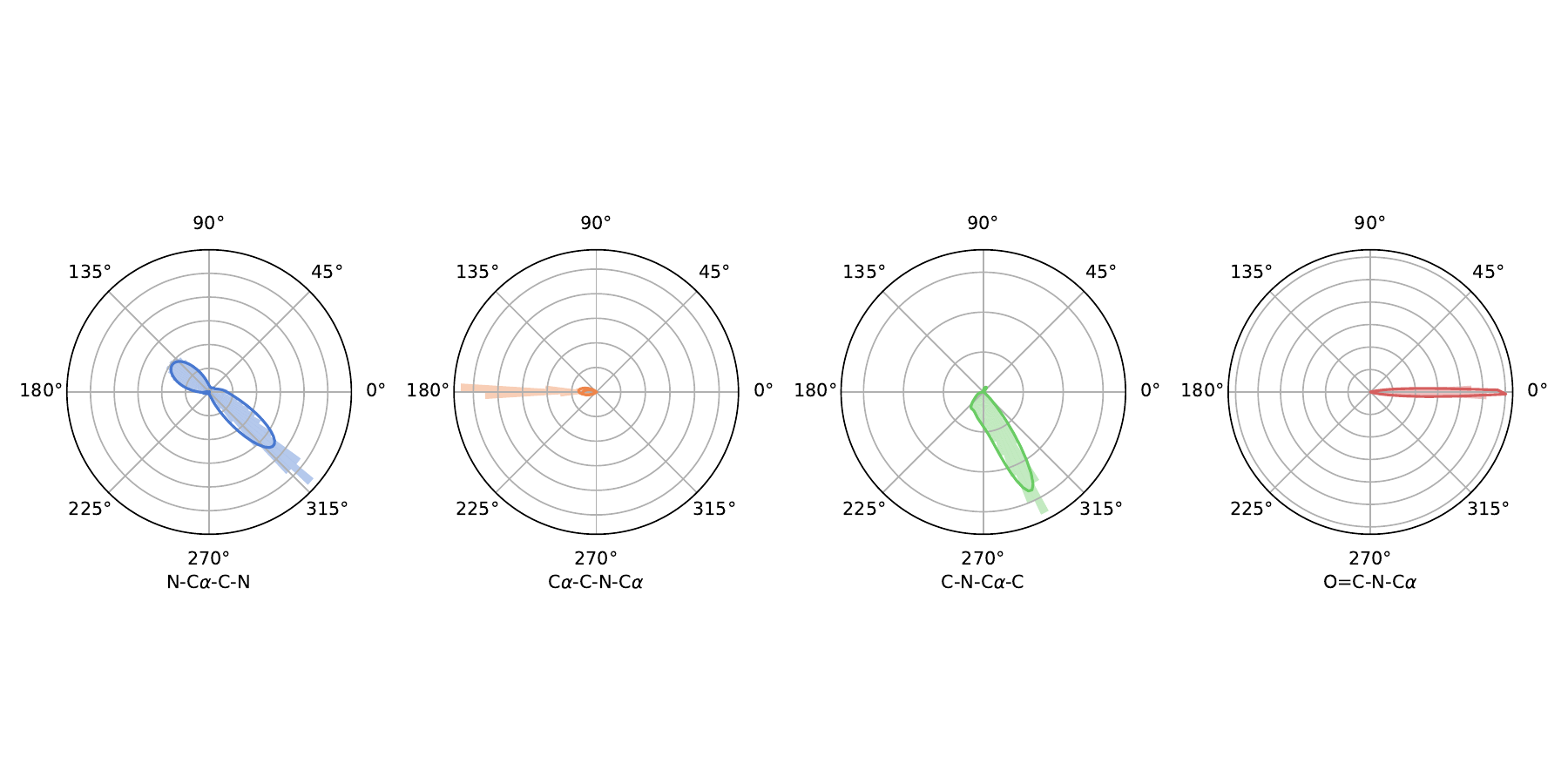}
    }\vspace{-0.5em}
    \caption{Distributions of flexible and inflexible geometries obtained by peptides in \texttt{PPBench2024} datasets.} \label{fig:geomdist}\vspace{-1em}
\end{figure}

Then, we give an empirical distribution on the peptide lengths of \texttt{PPBench2024}, in Figure.~\ref{fig:pplength}.

\subsection{Analysis on Geometry}
\label{app:geomanaly}
As the \textsc{PPFlow} models the internal redundant geometry of the peptides, here we give a statistical illustration to show the flexible geometries that need to be generated.
NeRF can use the following geometries in Figure.~\ref{fig:emplendist}, \ref{fig:empangdist} and \ref{fig:emptordist} to reconstruct the full backbone structures. For these internal geometries, it can be concluded that the torsion angles of `N-C$\alpha$-C-N' and `C-N-C$\alpha$-C' are the two most flexible geometries. Besides, `C$\alpha$-C-N-C$\alpha$' is theoretically inflexible since the constraints on peptide bond. However, in the observation, we find that it will deviate from the ideal value ($\pi$) a lot (about plus and minus $7^{\circ}$). In this way, we include it as another flexible geometries that the model needs to generate. These three torsion angles are named $\phi$, $\psi$ and $\omega$, respectively in formal definition.

\subsection{Experiment}
Here we give the hyper-parameters and other training details.
The learning rate $lr$ is $5e-5$. In all training, the max training iteration is $200000$. \texttt{LambdaLR} schedule is used, with \texttt{lr\_lambda} is set as $0.95\times lr$.  The batch size is set 16 or 32, because it affects the performance little.
In the neural networks, we set the MLP for extracting pair relations as 2 layers with hidden dimension as 64, and the MLP for single amino acid as 2 layers with hidden dimension as 128. Following, 6 layers of transformer are stacked behind, and the final layer is the LOCS which has been discussed before.
\end{document}